\documentclass[sigconf]{acmart}

\AtBeginDocument{%
  }

\copyrightyear{2025}
\acmYear{2025}
\setcopyright{cc}
\setcctype{by}
\acmConference[ISCA '25]{Proceedings of the 52nd Annual International Symposium on Computer Architecture}{June 21--25, 2025}{Tokyo, Japan}
\acmBooktitle{Proceedings of the 52nd Annual International Symposium on Computer Architecture (ISCA '25), June 21--25, 2025, Tokyo, Japan}
\acmDOI{10.1145/3695053.3731088}
\acmISBN{979-8-4007-1261-6/2025/06}

\usepackage{amssymb} 
\usepackage{pifont}  
\usepackage{multirow}
\usepackage{subcaption}
\usepackage{url}
\usepackage{listings}

\usepackage{multicol}
\usepackage{listings}
\usepackage{xcolor}
\definecolor{bg}{RGB}{250,250,250}
\lstdefinestyle{mystyle}{
  backgroundcolor=\color{bg},
  basicstyle=\ttfamily\scriptsize,
  frame=single,
  showstringspaces=false,
  tabsize=6,
  breaklines=false
}

\usepackage[absolute,overlay]{textpos}

\begin{document}

\title{Reconfigurable Stream Network Architecture}

\author{Chengyue Wang}
\orcid{0009-0006-7481-018X}
\affiliation{%
  \institution{University of California, Los Angeles}
  \city{Los Angeles}
  \state{California}
  \country{USA}
}
\email{chengyue@ucla.edu}

\author{Xiaofan Zhang}
\affiliation{%
  \institution{Google}
  \city{Mountain View}
  \state{California}
  \country{USA}}
\email{xiaofanz@google.com}

\author{Jason Cong}
\affiliation{%
  \institution{University of California, Los Angeles}
  \city{Los Angeles}
  \state{California}
  \country{USA}
}
\email{cong@cs.ucla.edu}

\author{James C. Hoe}
\affiliation{%
  \institution{MangoBoost and Carnegie Mellon University}
  \city{Pittsburgh}
  \state{Pennsylvania}
  \country{USA}
}
\email{jhoe@andrew.cmu.edu}

\renewcommand{\shortauthors}{Chengyue Wang et al.}

\begin{abstract}



As AI systems grow increasingly specialized and complex, managing hardware heterogeneity becomes a pressing challenge.
How can we efficiently coordinate and synchronize heterogeneous hardware resources to achieve high utilization?
How can we minimize the friction of transitioning between diverse computation phases, reducing costly stalls from initialization, pipeline setup, or drain?
Our insight is that a network abstraction at the ISA level naturally unifies heterogeneous resource orchestration and phase transitions.

This paper presents a Reconfigurable Stream Network Architecture (RSN), a novel ISA abstraction designed for the DNN domain.
RSN models the datapath as a circuit-switched network with stateful functional units as nodes and data streaming on the edges. Programming a computation corresponds to triggering a path.
Software is explicitly exposed to the compute and communication latency of each functional unit, enabling precise control over data movement for optimizations such as compute-communication overlap and layer fusion.
As nodes in a network naturally differ, the RSN abstraction can efficiently virtualize heterogeneous hardware resources by separating control from the data plane, enabling low instruction-level intervention.

We build a proof-of-concept design RSN-XNN on VCK190, a heterogeneous platform with FPGA fabric and AI engines. 
Compared to the SOTA solution on this platform, it reduces latency by 6.1x and improves throughput by 2.4x–3.2x.
Compared to the T4 GPU with the same FP32 performance, it matches latency with only 18\% of the memory bandwidth. 
Compared to the A100 GPU at the same 7nm process node, it achieves 2.1x higher energy efficiency in FP32.

\end{abstract}





\begin{textblock}{10}(12,0.4)
  \includegraphics[height=1.2cm]{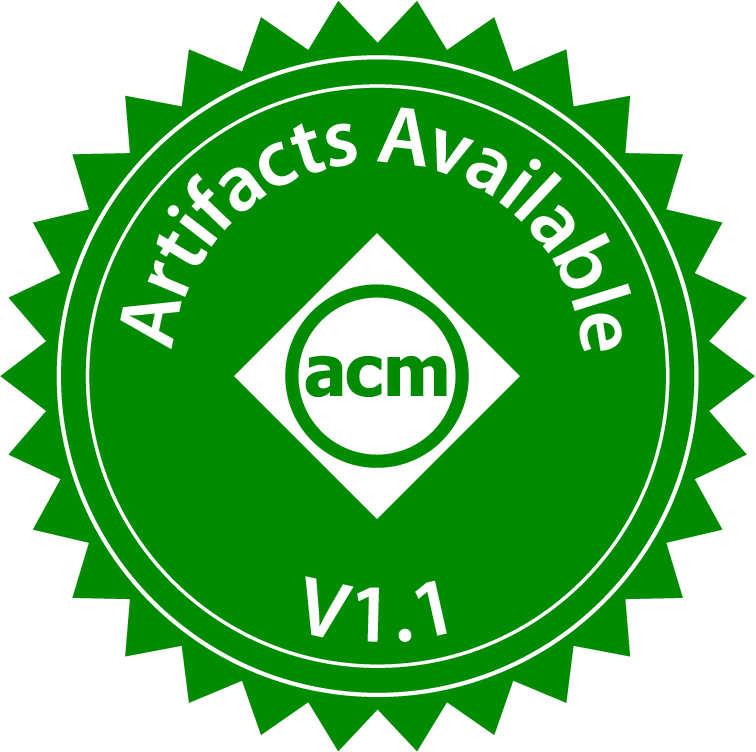}
  \hspace{0.0cm}
  \includegraphics[height=1.2cm]{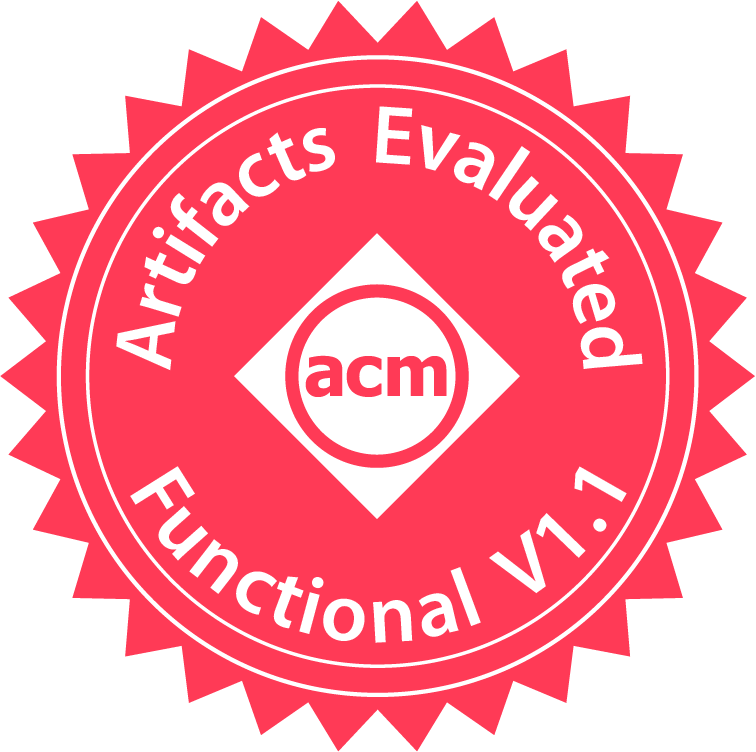}
  \hspace{0.0cm}
  \includegraphics[height=1.2cm]{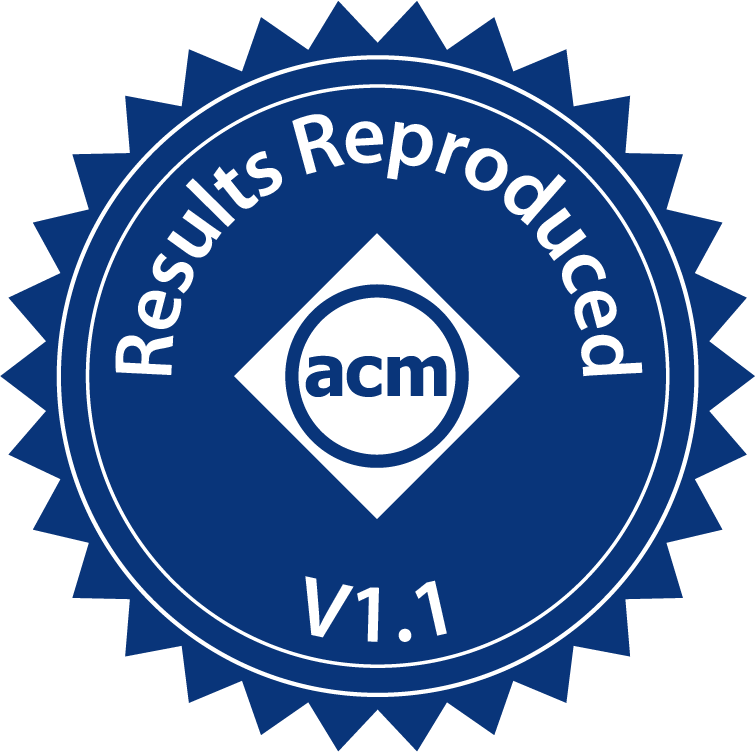}
\end{textblock}

\maketitle

\section{Introduction}
Artificial intelligence is driving dramatic demands for high performance and energy efficiency in computing.
While GPUs have seen great success in AI, some critics argue that their general-purpose support for the SPMD model introduces unnecessary overheads. This has motivated the emergence of ASIC-based AI accelerators such as TPU~\cite{tpu} and Groq~\cite{groq}.
Although ASICs offer good efficiency, they require lengthy chip development cycles. FPGA-based accelerators are being explored because they allow hardware datapaths to be reconfigured without the need to tape out a new chip.
The future of AI hardware lies in a tug-of-war between specialization and generalization. 
AMD’s Versal VCK190 exemplifies this trend as a heterogeneous platform that combines hardened AI engines (AIEs) with standard reconfigurable FPGA fabric, bringing together the efficiency of ASICs with the flexibility of FPGAs.

However, such heterogeneous platforms face fundamental challenges in resource orchestration. 
On the FPGA side, changing functionality over time causes significant stalls because it takes a sub-second latency to load a new bitstream.
Overlay techniques mitigate this by layering a virtual, reconfigurable architecture over the physical FPGA fabric so that functionality can change without reloading bitstreams ~\cite{fpga_overlay_book}.
However, current DNN overlays typically adopt coarse-grained, von Neumann-style instruction set architectures (ISAs) that control execution at the layer granularity ~\cite{overlay-dla, overlay-opu, brainwave, overlay-intel-npu, npe, overlay-dfx, overlay-flightllm, overlay-tvm, overlay-mangoboost-nlp, overlay-hybriddnn, overlay-graphagile, overlay-cnn2024, fet-opu-u280}. This coarse granularity severely restricts execution patterns, resulting in most overlays being only able to run one convolution or matrix multiplication (MM) layer at a time.
In contrast, the AIE part consists of 400 lightweight processors connected by a network-on-chip (NoC). Each core is driven by a fine‑grained instruction stream that operates on individual scalars or short vectors. Considerable effort is required to coordinate so many cores and keep them all busy. 
The huge differences in \emph{execution models, instruction interfaces}, and \emph{resource characteristics} within a heterogeneous system make it difficult to orchestrate resources across distinct components (spatially) and across the different execution patterns required by diverse DNN layers (temporally).

\begin{figure}  
    \centering
    \includegraphics[width=\linewidth]{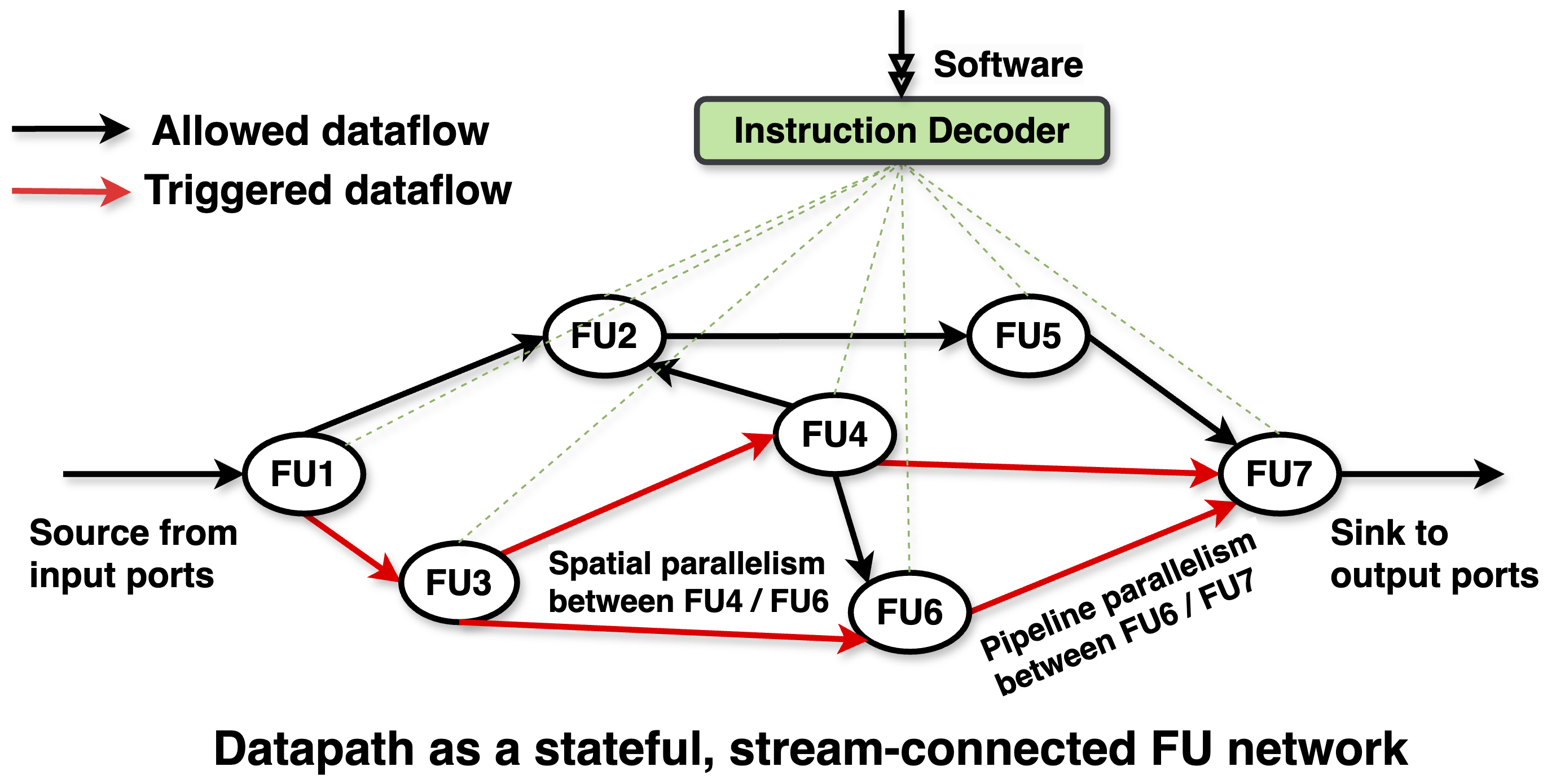}
     \vspace{-15pt}
    \caption{Reconfigurable Stream Network Overview}
    \label{fig:intro}
    \vspace{-10pt}
\end{figure}

Motivated by these observed challenges, we ask the question:
\emph{What is the right abstraction for bridging software with highly heterogeneous hardware?} 
Ideally, it should meet two key requirements:

\begin{itemize}
\item \textbf{Flexibility:} 
Computation and bandwidth must be flexibly allocated to support different phases, such as prolog, steady state, and epilog within a layer, as well as varying operator types and tensor shapes across layers. 
An ideal ISA should support low-cost phase initialization and give software fine-grained control over data movement to minimize pipeline stalls through techniques like overlapping and interleaving.
\item \textbf{Heterogeneity:} 
Coordinating components with different control paradigms requires an inclusive abstraction.
For example, while AIEs use a fine-grained instruction set, such ISA styles are too costly for FPGAs due to their lower operating frequency. The FPGA fabric itself also contains heterogeneous resources, such as BRAMs, DSPs, FFs, and LUTs. 
Meanwhile, the system experiences high volumes of data movement across components that are all synchronized at the nanosecond level. 
An ideal ISA should efficiently support this high-volume, heterogeneous parallelism.
\end{itemize}

Despite these challenges, DNNs present unique application opportunities. 
While they are compute- and memory-intensive, their execution patterns are highly repetitive and predictable.
DNNs' execution patterns have low information entropy, meaning that the total amount of information required to encode the controls of execution is small.
This observation suggests that if the architecture provides \emph{flexible instruction-to-data granularity}, bridging coarse layer-level and fine data-level granularity, the overall control overhead can remain low.
Furthermore, the deterministic nature of DNN execution allows for \emph{compile-time analysis} of data dependencies, eliminating the need for runtime discovery.

Our key insight is that introducing a network abstraction at the ISA level offers an elegant and unified solution to the challenges of coordinating heterogeneous resources and managing execution-phase transitions.
We propose a \textbf{reconfigurable stream network architecture}, where the datapath is abstracted as a specialized circuit-switched network of stateful functional units (FUs), as shown in Fig.~\ref{fig:intro}. 
Conceptually, programming a computation corresponds to triggering a circuit path in the network, 
with data sourced from input ports, streamed through FUs, and then sunk back to output ports. 
Multiple non-conflicting paths can be established to utilize available FUs for spatial parallelism between \emph{data-independent} computations. 
The output of one path can feed the input of another path for pipeline parallelism between \emph{data-dependent} task-level computations.
Instead of specifying the movement of every scalar or tensor, instructions control streams of data movement with high-level control information, enabling 1 byte of instruction to drive up to 1.6 GFLOPs of computation in our prototype.

The contributions of this work can be summarized as: 
\begin{itemize}
\item 
Propose the reconfigurable stream network architecture, detailing its principles and implementation.
\item
Identify the architectural bottlenecks in FPGA overlays
\item Develop the first design on FPGAs to achieve dynamic sequential linear layer pipelining and bandwidth orchestration. 
\item 
Prototype RSN-XNN on VCK190 and achieve a 6.1x latency reduction for BERT and a 2.4x-3.2x better throughput for BERT, VIT, NCF, and MLP compared to the state of the art.
\item 
Achieve the best GEMM implementation on VCK190. 
\item 
Present a quantitative comparison with T4, V100, A100, and L4 GPUs that shows the need for FPGAs to continue integrating ASICs for high performance and bandwidth.
\item 
Open source RSN-XNN to contribute to the community
\end{itemize}

\begin{figure}
    \centering
    \includegraphics[width=0.8\linewidth]{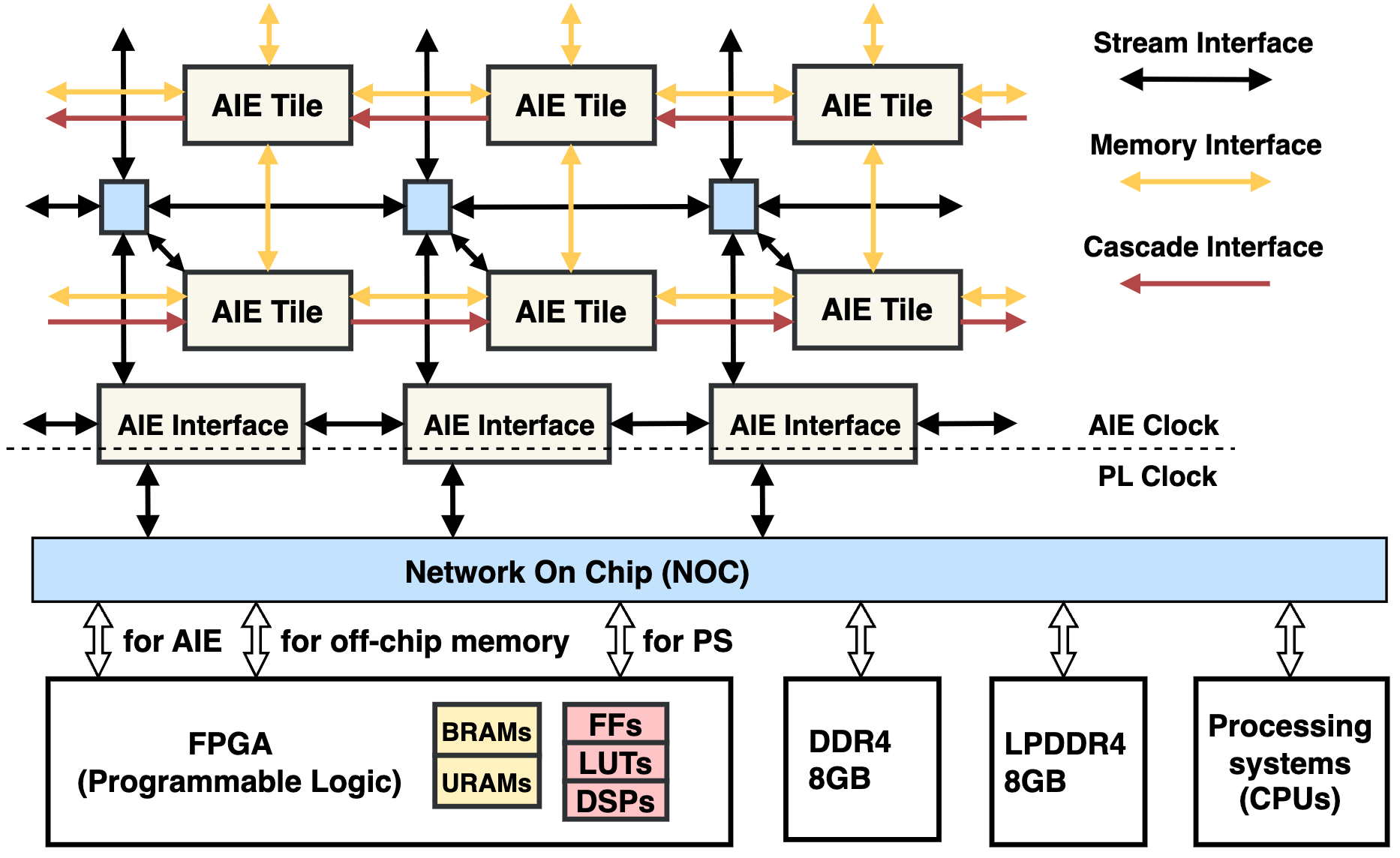}
    \caption{Versal ACAP VCK190 Block Diagram}
    \label{fig:vck190}
    \vspace{-10pt}
\end{figure}

\begin{table*}
\caption{Inter-Linear Layer Execution and Customization Flexibility Comparison of DNN Accelerators}
  \vspace{-7pt}
  \footnotesize{The 2nd to 6th columns target FPGA platforms, while the 7th column targets ASIC-based platforms. "Layer" here means linear operations like matrix multiplication and convolution.
}
\resizebox{\linewidth}{!}{
  \label{tab:hardware_capabilities}
  \footnotesize

  \begin{tabular}{p{5.9cm}p{1.7cm}p{0.5cm}p{1.2cm}p{1.4cm}p{1.2cm}p{2.2cm}p{1.2cm}}

    \toprule
    Supported Execution Features & NPU, etc. \newline\scriptsize	
   \cite{overlay-opu, brainwave, overlay-intel-npu, npe, overlay-dfx, overlay-flightllm, overlay-tvm, overlay-mangoboost-nlp, overlay-hybriddnn, overlay-graphagile, overlay-cnn2024, fet-opu-u280}  & DLA \cite{overlay-dla} & HPIPE, etc.\newline \cite{dnnbuilder, hpipe, llmfpga_cornel, via-u50, FTRANS}  & CHARM, etc.\newline \cite{charm, dnnexplorer} & TGPA, etc. \newline \scriptsize \cite{tgpa, SSR_2024, DeepBurning, multi-pu}   &  \scriptsize ASIC-based flexible \newline dataflow accelerators \cite{tangram, maestro_toppicks2020, set, scaledeep, atomic-dataflow, ISOSceles, cai2024gemini, SambaNova-paper} &  \textbf{RSN-\newline XNN (this work)} \\
    \midrule
    Software programmable & \checkmark & \checkmark & $\times$ & $\times$ & $\times$ & \checkmark & \textbf{\checkmark}\\ \hline
    Low instruction-level intervention & \checkmark & \checkmark & --- & --- & --- & $\times$ & \textbf{\checkmark}\\
    Remove redundant circuits in logic and interconnects & \checkmark & \checkmark & \checkmark & \checkmark & \checkmark & $\times$ & \textbf{\checkmark} \\
    Customize an FU for each layer at the bit level & $\times$ & $\times$ & \checkmark & $\times$ & $\times$ & $\times$ & \textbf{$\times$}\\
    Allocate the number of FUs based on layer shape & $\times$ & $\times$ & \checkmark & \checkmark & \checkmark & \checkmark & \textbf{\checkmark}\\ \hline
    Allocate all FUs for the same or mathematically fused layers (A, B, simplified C) & \checkmark & \checkmark & $\times$ & $\times$ & $\times$ & \checkmark & \textbf{\checkmark}\\
    Interleave dependent layers, one tile at a time (enhanced A) & $\times$ & \checkmark & $\times$ & $\times$ & $\times$ & \checkmark & \textbf{$\times$}  \\
    Spatially execute independent layers (C) & $\times$ & $\times$ & \checkmark & \checkmark & \checkmark & \checkmark & \textbf{\checkmark}\\    
    Spatially pipeline dependent layers (D) & $\times$ & $\times$ & \checkmark & $\times$ & \checkmark & \checkmark & \textbf{\checkmark}\\
    Dynamic chain of pipelined FUs (A, B, C, D)& $\times$ & $\times$ & $\times$ & $\times$ & $\times$ & \checkmark & \textbf{\checkmark}\\ \hline
    Overlap the prolog and epilog phases of layers & $\times$ & \checkmark & $\times$ & $\times$ & $\times$ & \checkmark & \textbf{\checkmark}\\
    Finely interleave off-chip load/store within the same layer & $\times$ & $\times$ & $\times$ & $\times$ & $\times$ & \checkmark & \textbf{\checkmark}\\
    \bottomrule
  \end{tabular}
}
\end{table*}

\section{Background and Motivation}
\subsection{AI Hardware and Versal Architecture}


The extreme computing needs of AI have led to AI-specialized hardware improvements such as Intel’s Xeon with vector neural network instructions~\cite{intel_cpu} and NVIDIA's A100/H100 with tensor cores~\cite{nvidia_a100, nvidia_h100}.
However, advocates for domain-specific hardware point out the unnecessary area and energy overheads in CPUs/GPUs due to general-purpose hardware features like cache and complex dependency controls~\cite{tpu}.
Google TPUs~\cite{tpu} and Groq AI processors~\cite{groq} simplify hardware by using software-managed scratchpads or communication for deterministic execution.
Unlike ASIC-based accelerators that require a lengthy chip development cycle, FPGA-based accelerators offer a faster "time-to-solution" because they can reconfigure datapaths without re-tapping. 
Traditional FPGAs have a low multiply-accumulate performance, which is mainly delivered by DSPs for traditional signal processing tasks~\cite{fpgasurvey}. To improve this, FPGA vendors are combining coarse-grained computing units with standard FPGA fabrics to exploit the high frequency, energy efficiency, and unit density of ASICs.
For example, AMD's Versal ACAP combines the FPGA fabric with a processor array~\cite{versal}, and Intel's Stratix NX integrates hardened tensor blocks directly into the FPGA fabric~\cite{stratix10}.

This work is demonstrated on the Versal VCK190 evaluation kit~\cite{vck190}, the first FPGA kit with AI engines. Fig.~\ref{fig:vck190}(a) depicts its system-level block diagram, which contains the processing system (PS), the programmable logic (PL), the AIE array, off-chip memories, and on-chip networks.
PS contains ARM CPUs and PL contains traditional FPGA resources such as LUTs, FFs, DSPs, and on-chip RAMs (4MB of BRAMs and 16MB of URAMs). 
The AIE array has 8 rows by 50 columns of AIE tiles, providing a peak throughput of 8 TFLOPS for FP32.
Each AIE tile has a 1.25 GHz 7-way VLIW processor and 32KB local software-managed scratchpad memory.
Moreover, the kit comes with one 8GB DDR4 and one 8GB LPDDR4, with a peak off-chip bandwidth of 25.6 GB/s and 32 GB/s.


\begin{figure}[t]
    \centering
    
    \includegraphics[width=\linewidth]{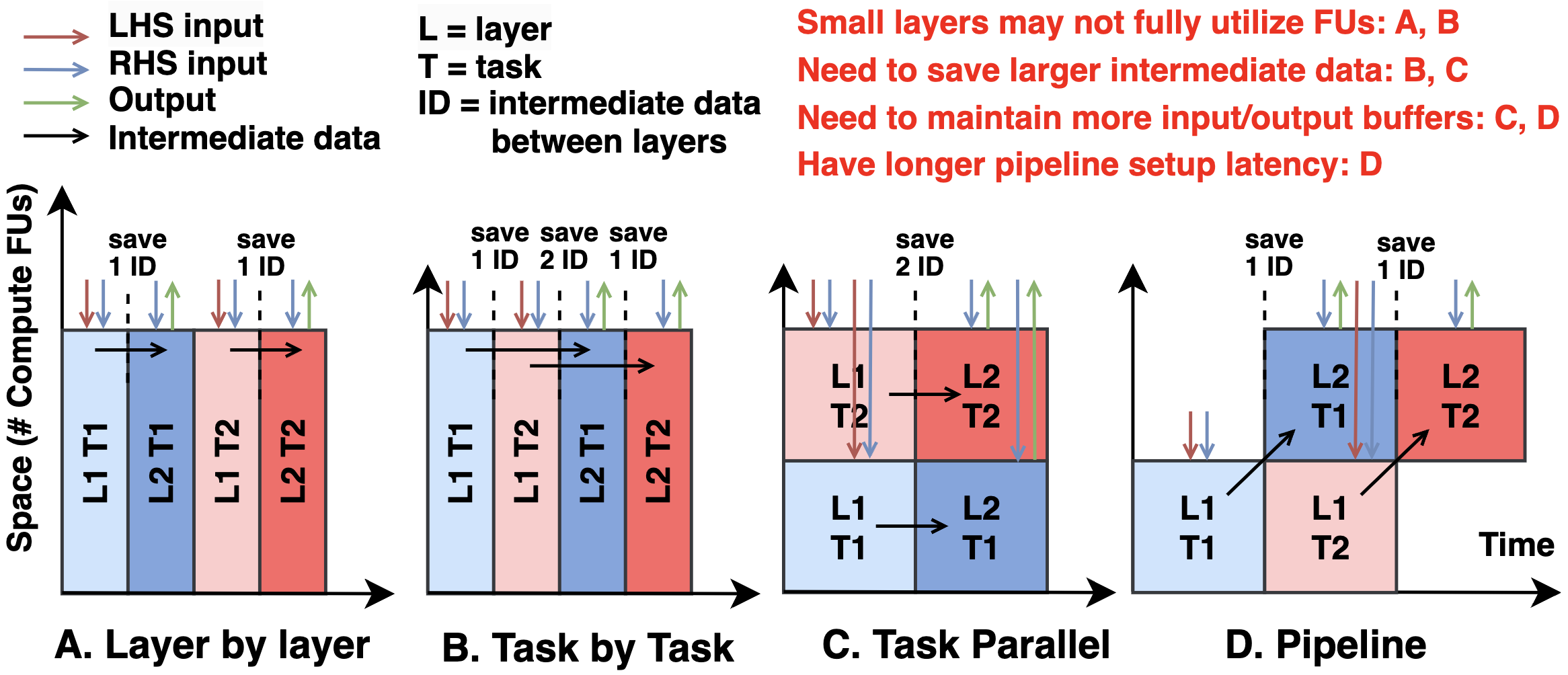}
    \vspace{-18pt} 
\caption{ Four Mapping Types and Their Disadvantages}
\vspace{-12pt}
    \label{fig:map4}
\end{figure}

\subsection{Inflexibility in Inter-Layer Execution for FPGA-Based DNN Accelerators}
\label{sec:background:interlayer}

Fig.~\ref{fig:map4} explains the trade-offs of different inter-layer mapping types using an example with two DNN tasks, each containing two dependent layers. 
Executing small layers one at a time, as in layer-by-layer (A) or task-by-task (B) mappings, can under-utilize compute units due to limited unrolling opportunities.
If the second layer only starts after the first layer is completed, intermediate data between the two layers must be stored, as in task-by-task (B) and task-parallel (C) mappings.
Due to limited on-chip memories, this often leads to large off-chip accesses.
The advantage of B over A is that the steady state in B is longer because the switching frequency between different layers is lower.
Spatially mapping two layers, like task-parallel (C) and pipeline (D) mappings, requires separate input or output buffers for the computations that occur at the same time.
If the right-hand side (RHS) input (e.g., weights across different batches) or the left-hand side (LHS) input (e.g., Q/K/V matrices in transformers) is shared between tasks, two smaller tasks can mathematically be fused into one larger task. 
This fusion is similar to type C, spatially mapping two small tasks to improve FU utilization but without the need for separate input/output buffers (simplified C).
Pipeline (D) mappings may cause longer delays due to the pipeline setup phases.

There are three common types of FPGA-based accelerators.
The \textbf{generic reusable FU} type uses a large, reusable FU to sequentially execute all layers~\cite{overlay-dla,overlay-opu, brainwave, overlay-intel-npu, npe, overlay-dfx, overlay-flightllm, overlay-tvm, overlay-mangoboost-nlp, overlay-hybriddnn, overlay-graphagile, overlay-cnn2024, overlay-xilinxdpu, fet-opu-u280, fpga-cnn, fpga-Caffeine, fpga-systolic, fpga-fpdnn, fpga-versatilesystolic, fpga-flexcnn,fpga-suhail}. This approach aligns with mapping types A and B, and can accommodate a simplified version of C when input operands are shareable. 
Intel's DLA~\cite{overlay-dla} supports an enhanced version of type A, processing one tile at a time. 
The \textbf{fully pipeline} type allocates a customized FU for each layer,
allowing direct forwarding of intermediate feature maps between pipelined FUs (type C, D) without writing them back to off-chip~\cite{dnnbuilder, hpipe, llmfpga_cornel, via-u50, FTRANS}. 
However, deep models may not fit and a longer end-to-end latency may result from pipeline setup. 
The \textbf{multiple reusable FUs} type balances between the previous two types by creating multiple customized FUs and strategically assigning layers to them.
DNNExplorer uses individual FUs for initial layers and a generic FU for later layers~\cite{dnnexplorer}, whereas CHARM uses separate FUs for large and small layers~\cite{charm} (type C). 
Other designs use a static chain of pipelined FUs to execute one segment of DNNs at a time, with SSR~\cite{SSR_2024} and~\cite{multi-pu} at a batch granularity, and TGPA~\cite{tgpa} and~\cite{DeepBurning} at a tile granularity (type C, D).
Static pipelined datapaths inherently lead to mismatches and reduced performance when faced with diverse layers.
The second and third columns in Table~\ref{tab:hardware_capabilities} reveal that current FPGA DNN overlays only adopt a generic reusable FU approach. 
Most of them execute one layer at a time~\cite{overlay-opu, brainwave, overlay-intel-npu, npe, overlay-dfx, overlay-flightllm, overlay-tvm, overlay-mangoboost-nlp, overlay-hybriddnn, overlay-graphagile, overlay-cnn2024, overlay-xilinxdpu, fet-opu-u280}, while DLA executes one tile at a time and features flexible stream buffers to prefetch layers~\cite{overlay-dla}. 
Fixed-function designs, as a category, offer greater overall inter-layer execution flexibility compared to overlays. However, each individual design exposes only a subset of that flexibility and rarely supports fine-grained off-chip bandwidth mapping~\cite{dnnbuilder, hpipe, llmfpga_cornel, charm, dnnexplorer, tgpa, SSR_2024, DeepBurning, multi-pu}.
ASIC-based flexible dataflow accelerators target tiled FUs interconnected via an NoC~\cite{tangram, maestro_toppicks2020, set, scaledeep, atomic-dataflow, ISOSceles, cai2024gemini, SambaNova-paper}. 
Most academic studies primarily focus on scheduling policies and rely on simulation-based experiments \cite{code-stanfordmastnn, code-maestroahw, code-github2024gemini, code-setisca2023core}, assuming FUs are controlled by a fine-grained ISA that can flexibly route data following the specified schedules.
SambaNova RDU ~\cite{SambaNova-paper} is an industry dataflow accelerator that supports aggressive operator fusion.
Compared to ASIC-based accelerators, RSN-XNN can intentionally exclude unnecessary features such as tile-level interleaving to save circuits in logic and interconnects. 
For applications requiring this feature, new RSN overlays can be built to support it.

Although FPGA-based studies offer better datapath customization, they have less flexibility in the use of compute and bandwidth resources compared to ASIC-based studies.
One key reason is current overlays suffer from serialization at the layer granularity, as discussed in Section \ref{sec:background:overlay}.  
RSN-XNN has the highest execution flexibility among FPGA designs, close to the levels typically assumed in ASIC studies. It is the first design on FPGAs that enables both dynamic layer fusion and fine-grained bandwidth mapping.

\subsection{Serialization in FPGA-Based DNN Overlay}
\label{sec:background:overlay}

There are two popular styles for building FPGA-based AI accelerators: fixed function 
\cite{dnnbuilder, hpipe, llmfpga_cornel, charm, dnnexplorer, tgpa, SSR_2024, DeepBurning, multi-pu} 
and overlay 
\cite{overlay-dla, overlay-opu, brainwave, overlay-intel-npu, npe, overlay-dfx, overlay-flightllm, overlay-tvm, overlay-mangoboost-nlp, overlay-hybriddnn, overlay-graphagile, overlay-cnn2024, overlay-xilinxdpu}. 
Fixed-function designs are efficient when flexible datapath reuse is unnecessary. 
However, it must integrate the reuse logic for every possible execution pattern directly into the datapaths, otherwise it suffers from sub-second bitstream reconfiguration latency. 
DNN overlays reintroduce the “stored program” concept into FPGAs' native dataflow execution by using instructions to control temporal datapath reuse.
They act as virtual, reconfigurable architectures that sit on top of the physical FPGA fabric so that functionality can change by changing instructions~\cite{fpga_overlay_book}.
Overlay users can compile and debug different DNN operations on the same FPGA bitstream in seconds and can also avoid hours or even days of generating a new bitstream.

Current DNN overlays primarily employ two styles of instruction set architectures: VLIW-like and RISC-like.
The VLIW-like style, as adopted in Intel's DLA~\cite{overlay-dla}, NPU~\cite{overlay-intel-npu}, and other works ~\cite{overlay-mangoboost-nlp, npe, overlay-cnn2024}, exploits massive parallelism by executing multiple FUs synchronously under a single wide instruction stream. 
For instance, Intel's NPU uses VLIW instructions of five macro-operations, each directing a different stage of the datapath including a matrix unit, a vector register file, two multi-function units, and a loader unit~\cite{overlay-intel-npu}. 
Alternatively, many works adopt a RISC-like instruction set, where every customized instruction maps to a straightforward operation~\cite{brainwave, overlay-flightllm, overlay-hybriddnn, overlay-graphagile, overlay-tvm, overlay-opu}. 
These operations can be divided into three categories: control (synchronization and scheduling), data movement (off-chip access), and computation (operations such as matmul, convolution, and activation).
For instance, Microsoft's Brainwave uses a single-threaded, in-order model~\cite{brainwave}, where a chain of dependent instructions, such as vector read/write and matrix-vector multiply, controls the execution of a single layer. 
While previous two ISA styles dominate, FGPU~\cite{overlay-fgpu} builds a GPU-like soft SIMT processor.
However, modern DNN overlays rarely adopt fine-grained, general-purpose ISAs, as they cannot match the performance of hardened GPUs without leveraging datapath customization.

The ISAs of the current FPGA-based DNN overlays are similar to those on a von Neumann model, 
in which a computer is logically composed of a central processing unit (containing an arithmetic logic unit and registers), a memory, and input/output interfaces. 
In this model, the arithmetic logic unit corresponds to the overlays' matrix engines and miscellaneous engines (typically for operations like pooling, ReLU, etc.), while registers correspond to the overlays' on-chip register files or buffers.
Similarly, the logical memory typically corresponds to off-chip device memory. 
To reduce control costs, current overlays predominately design coarse-grained instructions to control execution at the layer granularity~\cite{brainwave, overlay-intel-npu, npe, overlay-dfx, overlay-opu,  overlay-flightllm, overlay-tvm, overlay-mangoboost-nlp, overlay-hybriddnn, overlay-graphagile, overlay-cnn2024, overlay-xilinxdpu}.
However, since instructions are atomic, current overlays are inherently serialized at the layer granularity and are restricted in their execution patterns. 
This explains why Section~\ref{sec:background:interlayer} shows that they offer less inter-layer flexibility than FPGA-based fixed function designs.
In contrast, the RSN architecture conceptualizes the datapath as a network of stateful FUs, rather than keeping the program state in the logical memory or register files. 
Moreover, a single instruction can operate on flexible data granularity, ranging from the layer level for a low instruction cost to the multiple data level for precise control.

\subsection{Stall in Execution Phase Transition}

DNN overlays typically finish a layer by draining the entire pipeline before launching the next one, leaving compute units idle during the transition to the next execution phase. 
Although some add double‑buffering \cite{overlay-dfx} or prefetching \cite{overlay-dla} to hide part of this latency, none offers fine‑grained interleaving of memory accesses across phase boundaries (see Section \ref{section:bandwidthmap}).
This is because VLIW- or RISC-like ISAs treat each instruction as architecturally atomic, and an instruction must finish draining its results before executing the next one. 
Microarchitecturally, the drain of one instruction could overlap with the next one by cracking coarse-grained instructions into micro‑ops and having structures like store buffers; however, (1) the extra hazard tracking logic would add hardware complexity, (2) microarchitectures would still lack application-level knowledge (e.g., where the next instruction’s load gaps occur) to schedule a truly fine‑grained interleave, and (3) computation and communication would still interfere with each other when making overlap.

Instead, an execution phase in RSN is expressed as a decomposable path rather than a single atomic instruction.
Once the load and compute segments of that path finish, the control plane can immediately retarget them to the next phase while the store segment keeps draining results to off‑chip.  
RSN software can orchestrate the FU responsible for off-chip transfers to drain on-chip results during the next phase’s load gaps (see Section \ref{section:bandwidthmap}),
precisely controlling data movement to reduce stalls during phase transition. 
For example, RSN-XNN can split a 768K element output tile into 12 64K blocks and drain each block during load gaps between two 96K input loads of the next output tile, finely orchestrating 1 DDR channel on the board. 
Moreover, RSN can initiate a new phase with minimal instruction-level intervention. Only FUs with updated dataflow require new instructions, and since instructions carry only control information rather than data, they remain off the critical path.

\subsection{Virtualize Coarse-Grained Heterogeneity}
\label{sec:background:dataflow-vector}

The serialization bottlenecks of the von Neumann model have been extensively studied within the CGRA and dataflow architecture communities.
However, we find that current studies generally have limited support for coarse-grained heterogeneity \cite{dataflow-decoupledaccess, dataflow-triggered, programmable-stream-processor, stream-dataflow, MozartReuse, DSAGEN, cgra-heter-1, cgra-heter-2, cgra-heter-3, cgra-heter-4, cgra-heter-5, cgra-heter-6, cgra-heter-7, cgra-heter-8, cgra-homo-1, cgra-homo-2, cgra-homo-3, cgra-homo-4, cgra-homo-5, cgra-homo-6, cgra-AHA, cgra-Heterogeneous-ml, dmt-cgra, mt-cgra-1, mt-cgra-2, cgra-inductivemm}, 
which hinders their direct adoption in DNN overlays. 
First, CGRAs typically feature small, relatively uniform functional units, ranging from directly wired adders/multipliers to instruction-programmed processors. 
In contrast, FPGA-based DNN overlays have fewer, but significantly larger and more heterogeneous FUs.
For example, NPU \cite{overlay-intel-npu} includes a high-throughput matrix unit that delivers 10 INT8 TOPS, far exceeding the typical FU granularity in CGRAs. Additionally, the entire datapath contains only four more FUs—a vector register file, two multi-function units, and a loader—all very coarse-grained and highly functionally different.
This coarse-grained heterogeneity contrasts sharply with current heterogeneous CGRAs \cite{cgra-heter-1, cgra-heter-2, cgra-heter-3, cgra-heter-4, cgra-heter-5, cgra-heter-6, cgra-heter-7, cgra-heter-8}, which generally only vary in arithmetic or memory configurations.
Second, in FPGA-based DNN overlays, communications between different FUs involve much larger data volumes and are highly customized for the application. 
Compared to CGRAs (the majority uses a 16/32-bit datapath \cite{cgra-survey}), MeshB FU in our RSN-XNN must route 9K bits of data in one cycle (300 GB/s). This high-volume data movement is fundamentally different from small-scale, fine-grained communications typically targeted by CGRAs.

Studies have attempted to use CGRA architectures to virtualize FPGA resources; however, they generally exhibit the same limited coarse-grained heterogeneity we observed. Most designs \cite{overgen, cgra-fpga-homo-1, cgra-fpga-homo-2, cgra-fpga-homo-3} virtualize FPGA resources as homogeneous FU tiles, whereas CGRA-ME \cite{cgra-fpga-hete-1} offers minimal heterogeneity by allowing FUs to select from two sets of arithmetic operations. 

RSN virtualizes hardware resources into coarse-grained, heterogeneous FUs, enabling flexible high-volume data movement. 
By allowing FUs to be heterogeneous, RSN naturally maps to heterogeneous hardware resources. 
As long as resources can respond to the control plane and support sufficient stream execution to interact with other FUs, they can be virtualized using the FU abstraction. For example, AIEs can be virtualized as FUs via microprograms that coordinate their streaming behaviors with FPGAs, while the RSN programs are agnostic to whether a given FU is implemented on AIEs or FPGAs.

\section{Reconfigurable Stream Network Architecture}

\subsection{Abstraction}
\label{rsn-abstraction}
\begin{figure}[bt]
    \centering
    \includegraphics[width=1\linewidth]
    {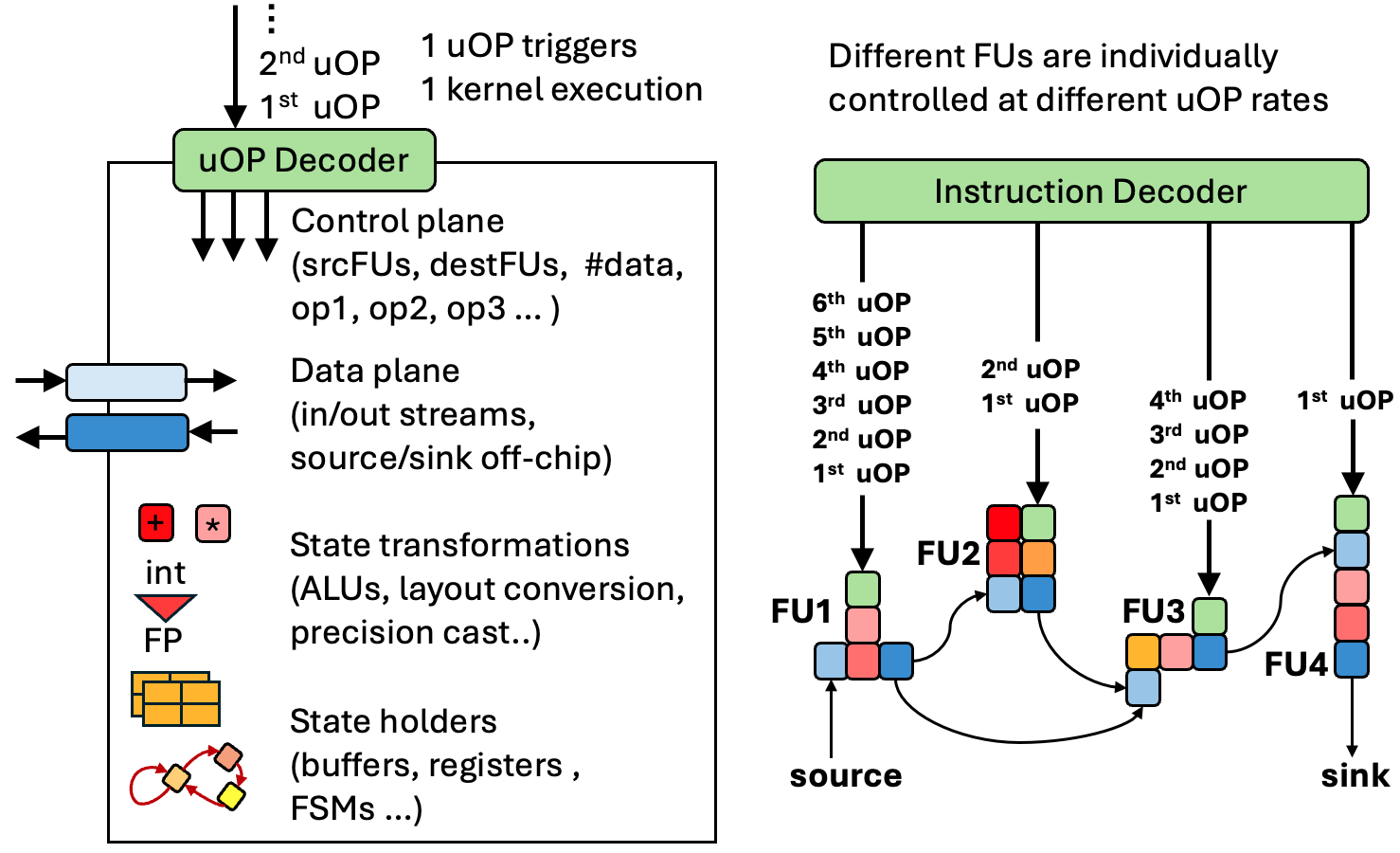}
    \vspace{-12pt}
    \caption{Functional Unit and Datapath Abstraction}
    \label{fig:abstraction}
    \vspace{-15pt}
\end{figure}

\textbf{Components:} An RSN computer consists of a datapath and an instruction decoder, with the datapath abstracted as a specialized circuit-switched network of stateful FUs. 
The data is sourced from off-chip, streamed and transformed through FUs, and ultimately sunk back off-chip.
An FU comprises a micro-operation (uOP) decoder, input and output ports, and customized modules designed to transform and hold states (Fig. \ref{fig:abstraction}).
A uOP decoder is the interface that makes an FU controllable by software according to the uOP sequence received.
This uOP sequence directs the method and the amount of data to be transformed and communicated (control plane). 
Ports include streams used for data communication between nodes, allowing the transmission of a continuous sequence of data from one source FU to another destination FU (data plane). 
This communication is latency-insensitive, meaning that the correctness of execution does not depend on timing, and the FUs are stallable.
Off-chip memories usually serve as the source and sink for the datapath through memory abstractions.  
Modules customized with state transformations (e.g., arithmetic operations and layout transformations) manipulate and process data, while state holders (e.g., buffers, registers, and FSMs) preserve both data and the associated execution states.

\begin{figure}[t]
\centering
\setlength{\columnsep}{10pt} 
\begin{multicols}{2}
\begin{lstlisting}[style=mystyle]
// GPU program
__global__ void GEMM(...) {
// Each thread computes 1 tile
}
GEMM<<<gridDim, blockDim>>>(...)
\end{lstlisting}
\columnbreak
\begin{lstlisting}[style=mystyle]
// RSN program
FU_LHS CtrlA // Load matrix A
FU_RHS CtrlB // Load matrix B
FU_Compute CtrlC // Perform MM
FU_OUT CtrlD // Store result
\end{lstlisting}
\end{multicols}
\vspace{-17pt}
\caption{Comparison between GPU's thread-based and RSN's stream-based programming models.}
\label{fig:gpu-vs-rsn}
\end{figure}

\textbf{Programming Model:} 
A computation in RSN is programmed by triggering a circuit path through the FU network. Each FU executes a sequence of kernels, with each kernel representing \emph{an atomic step in transforming the FU’s internal state}.
 Each uOP launches a single execution of the kernel, providing control information such as what transformation to perform, where to stream data to or from, and the length of each stream.
This notion of a kernel aligns with prior work on streaming processors \cite{Stream-proga, programmable-stream-processor, imagine-stream}, where a kernel represents a stream-based function that can be encapsulated as a custom instruction.
Unlike CUDA or OpenCL, where a kernel denotes a large, data-parallel function launched across many threads, RSN kernels are fine-grained operations, and the smallest units of scheduling and execution. 
Fig. \ref{fig:gpu-vs-rsn} compares a GPU program that expresses parallelism through threads and an RSN program that expresses parallelism through streams among FUs.
In GPU programming (left), many threads are launched to run the same kernel (GEMM) independently on different pieces of data.
In RSN programming (right), a computation is composed of instructions that launch the kernels on the involved FUs, with each CtrlX directing the behavior of its respective kernel. 
Parallelism in RSN arises not from replicated execution contexts but from activating different parts of the hardware to operate on streaming data in parallel. 
This makes RSN a more explicit and spatial model, enabling software to precisely control each stage of data movement.
For every piece of data sent by an FU’s output port, the destination FU must have a receive operation at its corresponding input port. 
The programmer must ensure that the number of sends from the producer kernel exactly matches the number of receives in the consumer kernels. If the sends are fewer than the receives, the receiving kernel will block indefinitely; if the sends exceed the receives, the producer kernel will block once the stream channel is full.

\textbf{Execution Model:} 
The RSN execution model closely aligns with its programming model, enabling intuitive mapping from software descriptions to hardware behavior. Each FU in the network maintains its own sequence of uOPs and can execute only one kernel at a time. Once a kernel completes, the FU fetches the next uOP from its local queue and stalls if no further instructions are available. Between FUs, execution is coordinated through latency-insensitive streams, which decouple data movement from timing and support stallable communication. This decentralized, stream-based execution allows independent FUs to operate in parallel and at different uOP rates. Efficient task-level pipelining can be achieved by chaining the output ports of one FU to the input ports of another.

\begin{figure}[t]
    \centering
    \includegraphics[width=\linewidth]
    {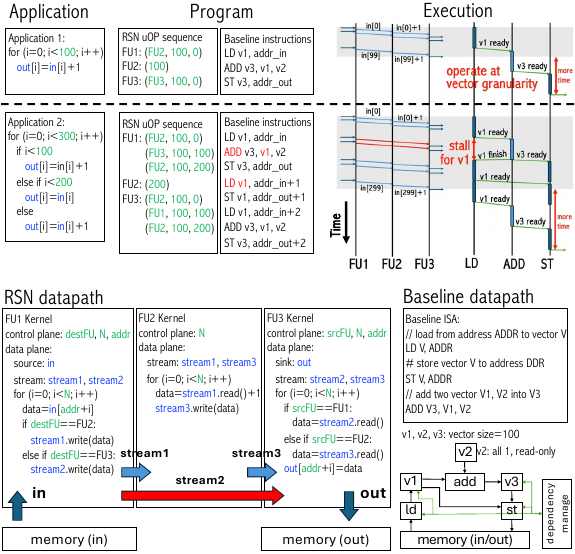}
    \caption{Comparison of RSN and Baseline Datapath}
    \label{fig:fumapping}
    \vspace{-10pt}
\end{figure}

\textbf{Simple Overlay:} 
Fig.~\ref{fig:fumapping} presents a comparison between RSN-based implementations and a baseline approach that employs a RISC-like ISA \cite{brainwave, overlay-flightllm} to virtualize the datapath for the same two applications.
The RSN datapath comprises three FUs with three streams: FU1$\rightarrow$FU2, FU1$\rightarrow$FU3, and FU2$\rightarrow$FU3. FU1 can read \(N\) data from the source \emph{in} at address \emph{addr} and write to FU2 or FU3. 
FU2 can increment data from FU1 by 1 and forward it to FU3. FU3 can store \(N\) data from FU1 or FU2 to the sink \emph{out} at \emph{addr}.
The baseline overlay utilizes a vector ISA, having a datapath with 3 100-element vector registers and add/load/store units. 
Like most previous overlays, the connections between these units are customized for applications, rather than allowing all-to-all register connections. 
Application 1 increments 100 data elements from the source \emph{in} by 1 and then stores them in the sink \emph{out}. 
One single uOP commands one FU.
For example, FU1's uOP passes (FU2, 100, 0) to the control plane to direct FU1 to read 100 data from index 0 and forward them to FU2. 
In contrast, the baseline instructions use vector load/add/store instructions to achieve the same function. 
In program order, the add must start after the load is complete in order to obey the true dependency.
Application 2 only increments data at index 0-99 and 200-299, while directly copying data at index 100-199.
In the RSN uOP sequence, by increasing N from 100 to 200, FU2 still only needs one uOP, while FU1/FU3 need three uOPs to reconfigure the destination/source FUs.
The triggered paths can be reprogrammed flexibly to accommodate different applications. 
However, the baseline execution stalls due to WAR dependencies between instructions. 
The second load must wait for the first add to complete, since both instructions use register v1.

This type of data hazard is typically resolved in CPUs through register renaming. However, FPGA-based DNN overlays rarely apply this technique because their coarse-grained ISAs map "registers” to large on-chip memory, which makes it costly. 
Another solution is to introduce an additional load register to allow the program to explicitly specify double buffering.
However, this introduces redundant buffering states in cases where the application has no actual buffering needs.
In contrast, the RSN datapath eliminates the need for explicit buffering by exposing streams between FUs, enabling direct dataflow without intermediate register buffering.



\begin{figure}[t]
    \centering
    \begin{subfigure}[b]{\linewidth}
        \centering
        \includegraphics[width=0.88\linewidth]{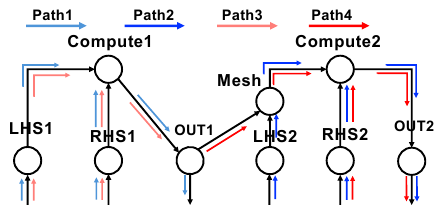}
        \caption{Datapath}
        \label{fig:datapath-2layer}
    \end{subfigure}
    \begin{subfigure}[b]{\linewidth}
        \centering
        \vspace{8pt}
        \includegraphics[width=\linewidth]{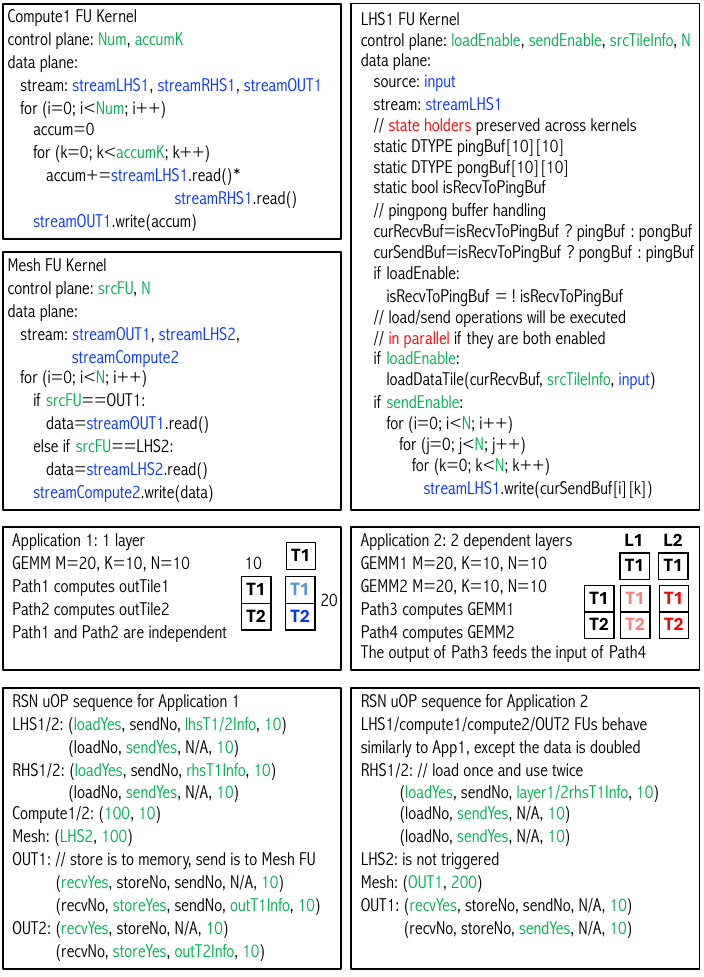}
        \caption{Functional Units, Applications and Programs}
        \label{fig:pipelined}
    \end{subfigure}
    
    \caption{A Flexible Datapath Supporting Dynamic Two Sequential Layer Pipelining.}
    \label{fig:combined-pipelining}
    \vspace{-10pt}
\end{figure}

\textbf{DNN Scenario:} 
With the network abstraction, multiple paths can be triggered in the datapath at the same time. Multiple independent paths can be established to exploit spatial parallelism by triggering separate FUs for data-independent computations, while the output of one path can feed into another path to allow for pipeline parallelism in data-dependent computations. 
Fig.~\ref{fig:datapath-2layer} exemplifies a flexible datapath that can either use all compute resources to execute a single 20x10x10 GEMM layer or pipeline two sequential 20x10x10 GEMM layers without sending intermediate data off-chip.  
If mapping one layer at a time, two independent paths are triggered: Path 1 uses Compute1, and Path 2 uses Compute2 FU. When mapping two pipelined layers, Path 3 executes layer 1 by fetching LHS and RHS input from the off-chip and stops after saving the output of layer 1 into OUT1 FU. Subsequently, Path 4 uses the output of Path 3 as the RHS input for layer 2 and Compute2 FU for computation.

In Fig.~\ref{fig:pipelined}, we provide the kernels of three FUs to help the reader better understand how RSN applies to DNN applications. 
LHS1 FU has two 10x10 buffers and one ping-pong buffer flag that can be preserved between kernels. The kernel first performs a pingpong buffer check. One buffer is used to save data loaded from the off-chip, and the other buffer is used to send data to Compute1 FU. 
If both load and send are enabled, they will be executed in parallel to overlap computation and data loading.
Since the buffers in LHS1 FU are large enough to store the entire K dimension, all data along the K dimension can be streamed continuously to Compute1 FU. 
For each of the Num iterations, Compute1 FU sums the products of accumK pairs from streamLHS1 and streamRHS1, then writes the completed result to streamOUT1.
Mesh FU routes data from either streamOUT1 or streamLHS2 to streamCompute2, based on the srcFU, ensuring the correct data flow for N iterations.

To program the datapath for Application 1 in Fig.~\ref{fig:pipelined}, we assign both Compute1/2 FUs to compute a single 10×10 output tile. 
Their control planes are configured to produce 100 outputs, each resulting from 10 accumulations. 
Mesh FU is configured to enable data flow along Path2.
For the input FUs, LHS and RHS, each reads a data tile in one uOP and sends it out in the subsequent uOP. Finally, OUT1/2 FUs receive data tiles and store them with two uOPs. 
Application 2 is programmed in a similar way.
This example shows how to program different DNN applications using different execution patterns.
Also, we can observe that the behavior of Compute1/2 remains consistent regardless of the mapping styles. 
This shows that we can just partially reprogram the datapath when switching between mapping styles, simplifying instruction issuing and decoding costs.


\subsection{Advantages of RSN Abstraction in DNN}

\textbf{Flexible execution patterns:}
To execute a DNN model, each layer typically involves distinct phases, such as prolog, steady-state, and epilog. When the steady-state is small, precise control over datapath behavior during the non-steady states becomes crucial. 
Across layers, the diversity in operation types and shapes necessitates flexible inter-layer computation resource mapping. 
The RSN abstraction enables precise control by exposing each FU’s properties to software and supporting flexible reprogrammable datapaths.

\textbf{Inherent parallelism:}
The DNN models have massive concurrency. 
In the von Neumann architecture, program states are stored in logical memory and register files, with instructions executed sequentially. 
This inherently limits parallelism due to the limited number of ports in the memory and register files. 
In contrast, as states are maintained within the FUs and distributed across the FU network, the number of data ports in an RSN datapath scales with the parallelism it can provide.


\textbf{Heterogeneity and customization:}
Modern hardware systems have heterogeneous resources.
Abstracting the actions that an FU can perform through kernels provides a unified way to manage this diversity and isolates the actual FU microarchitectures from the software. 
Modifications to an FU's actions impact only its neighboring FUs, preserving the integrity of other connections.
Moreover, abstracting the datapath as a network of FUs makes it composable, allowing each component to be independently optimized for customized tasks.



\textbf{Low instruction cost:}
Although efficient DNN execution requires precise control, the control for one DNN is regular and repetitive both within a layer and at the granularity of layers. 
The RSN architecture leverages this property to reduce control costs by supporting flexible instruction to data granularity and partial path reprogramming, as different FUs and compute phases demand different levels of control effort.
Additionally, data are locally synchronized between producer and consumer FUs through streams at the edges. This avoids centralized data dependency management, as data is not carried by the arrival of instructions.

\textbf{Determinism:}
The RSN execution model does not support prediction or speculative execution. While these features could potentially be added in the future, we justify the current choice by noting that DNN executions are generally deterministic.
The order of execution and data dependencies is known at compile time, and there is no need for runtime discovery. 
Runtime uncertainties in the system, such as variable DRAM latency and NoC transfer latency, are handled by latency-insensitive streaming protocols.

\begin{figure}[t]
    \centering
    \includegraphics[width=\linewidth]{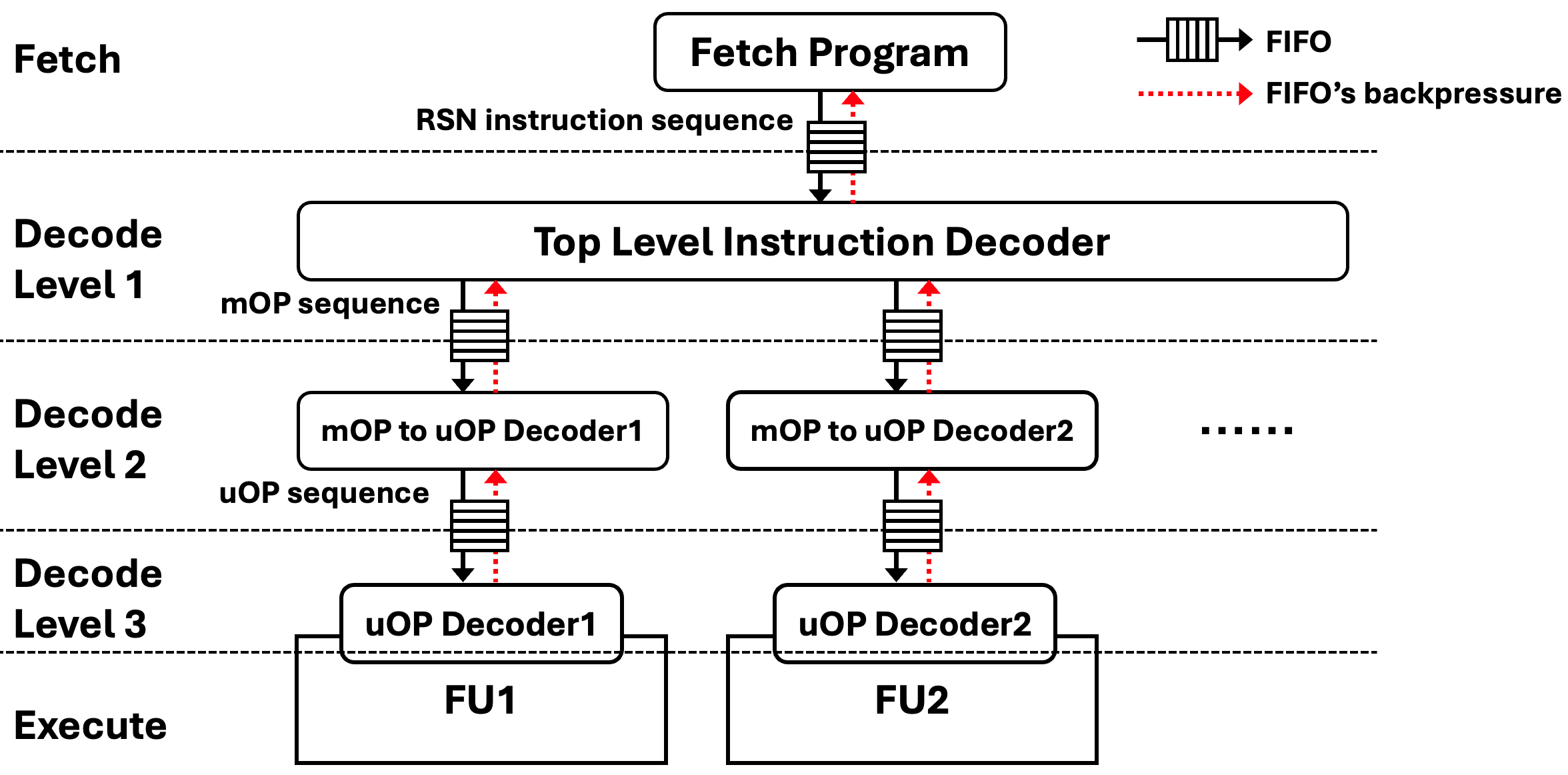}
    \caption{Instruction Decoder: Fuse uOP Streams into 1 RSN Instruction Stream}
    \label{fig:executionpipline}
    \vspace{-5pt}
\end{figure} 

\begin{figure}[t]
    \centering
    \includegraphics[width=1\linewidth]{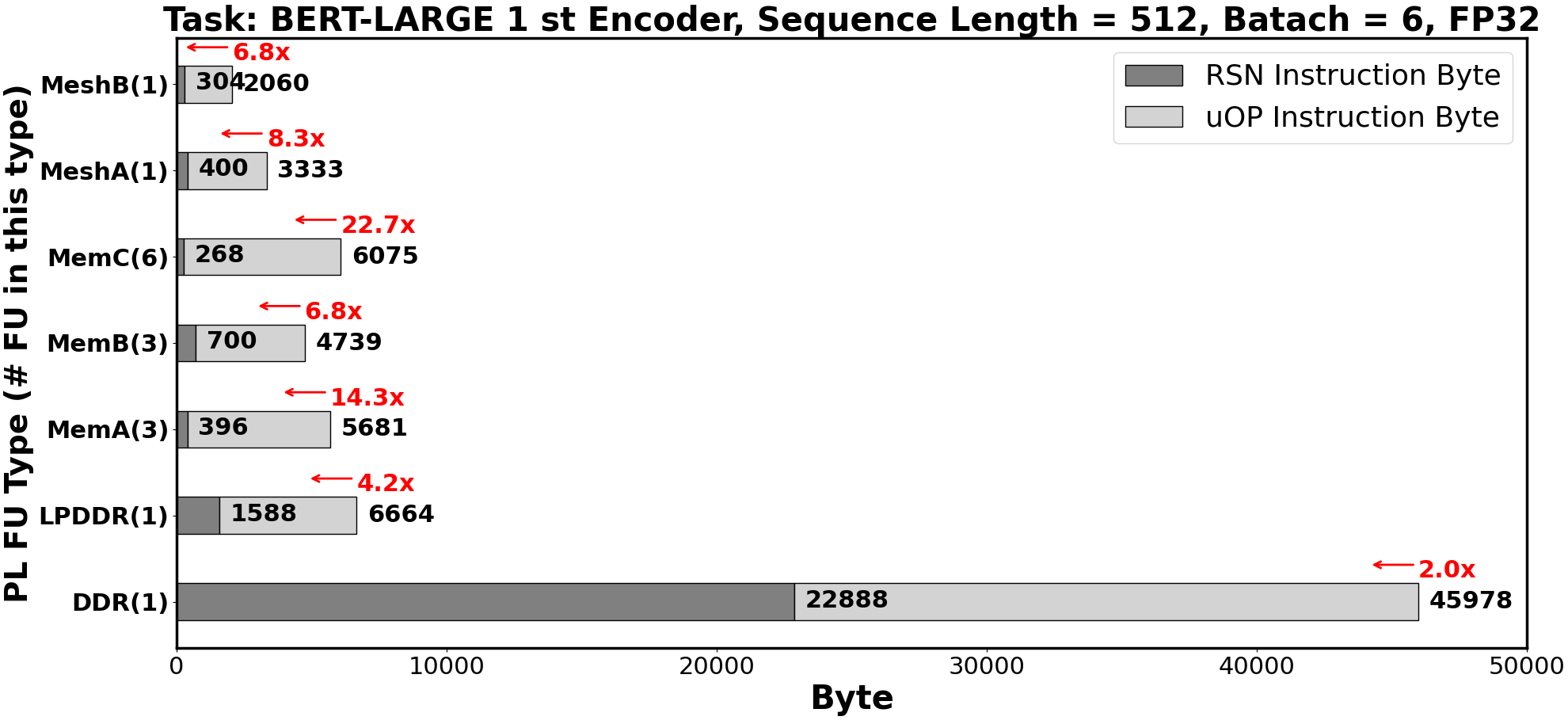}
    \vspace{-20pt}
\caption{RSN vs Translated uOP Size for Different FU Types}
\vspace{-10pt}
    \label{fig:instructionbyte}
\end{figure}

\subsection{Instruction Decoder}
While the concept of providing one instruction stream to one FU is straightforward, it suffers from the drawbacks of duplicated instruction fetch/decode units and increased programming complexity.
Similar to the solutions provided in the decoupled access/execute architecture \cite{dataflow-decoupledaccess}, we address this issue by physically merging multiple instruction streams into a single stream. 
As illustrated in Fig. \ref{fig:executionpipline}, the program is stored as a single sequence of RSN instructions, and the instruction decoder issues uOP sequences to the FUs.
Instead of merely interleaving different uOP sequences into a single RSN sequence, we introduce an intermediate level of decoding to enable RSN instruction reuse and enhance code efficiency.
For a full explanation, \textbf{Top-level decoder} receives an RSN instruction sequence consisting of UDP-like instruction packets, each with a 32-bit header and a payload section.  
The header includes:  
1) {\em opcode}: FU type;  
2) {\em mask}: selects targeted FUs;  
3) {\em last}: signals FU exit;  
4) {\em window size}: the number of RSN instructions in this packet;  
5) {\em reuse}: how many times this packet will be reused.  
The decoder converts the payload into macro-operations (mOPs) and sends them to the second-level decoders, with destinations specified by {\em opcode} and {\em mask}, and the number of mOPs specified by {\em window size}.  
\textbf{Second-level decoders} enable instruction packet reuse to improve code efficiency.  
A decoder first looks at the header to determine the {\em window size} and {\em reuse} of an instruction packet.  
It then processes the specified {\em window size} of mOPs, converts them into uOPs, stores them locally, and forwards them to the third-level decoder repetitively for the specified {\em reuse} count.  
This mechanism addresses common scenarios where a small sequence of uOPs is repeated, such as when an FU needs to send data to FU1 and then FU2, repeating the process 128 times.  
In this case, an instruction packet with a {\em window size} of 2 and a {\em reuse} of 128 can be created.  
FPGA's reconfigurability allows further customization. For example, we add {\em stride size} and {\em stride count} to some FUs to support strided off-chip accesses.  
\textbf{Third-level decoders} are attached to FUs and translate uOPs to control kernel execution. 

\textbf{Decoding efficiency:} 
Fig.~\ref{fig:instructionbyte} compares the size of RSN instructions needed to execute one BERT-Large encoder in our design with the size of the translated uOPs, categorized by FU types.  
This figure shows the diversity of controls required for different FU types.  
In terms of uOP size, FUs (LPDDR, DDR) that interact with off-chip memory require more controls than FUs (MeshA/B, MemA/B/C) that transform data on-chip through streaming interfaces.  
Moreover, the control of the DDR FU is more complex than that of the LPDDR FU because we enable complex interleaving of load/store feature maps in DDR, while LPDDR is only used to load read-only weights and biases.  
For the compression ratio of the RSN instruction to uOPs, the LPDDR FU and DDR FU have relatively lower ratios ($2 \sim 4.2$x) compared to other FUs ($6.8 \sim 22.7$x) because it is more difficult to exploit common patterns when addressing off-chip memories than when accessing stream interfaces.

\textbf{Deadlock:}
A decoder is back-pressured if its downstream FIFO is full.
The fetch unit issues instructions continuously until it encounters a stall. 
For instance, when FU1 is executing and waiting for FU2 to consume data, the fetch unit can stall because FU1 is unable to accept new uOPs. 
It remains stalled until FU1 completes its current kernels and consumes the next uOP.
In such cases, a deadlock may occur if the fetch unit stalls before fetching the instruction that directs FU2 to consume the data from FU1.
To address this, increasing the FIFO depth in the decoding system can help fetch units continue to retrieve subsequent instructions. While comprehensive deadlock prevention is more complex and beyond the scope of this paper, we report that setting FIFO depths to six between uOP and mOP decoders is deadlock-free in our implementation.

\begin{figure}[t]
    \centering
    \includegraphics[width=1\linewidth]{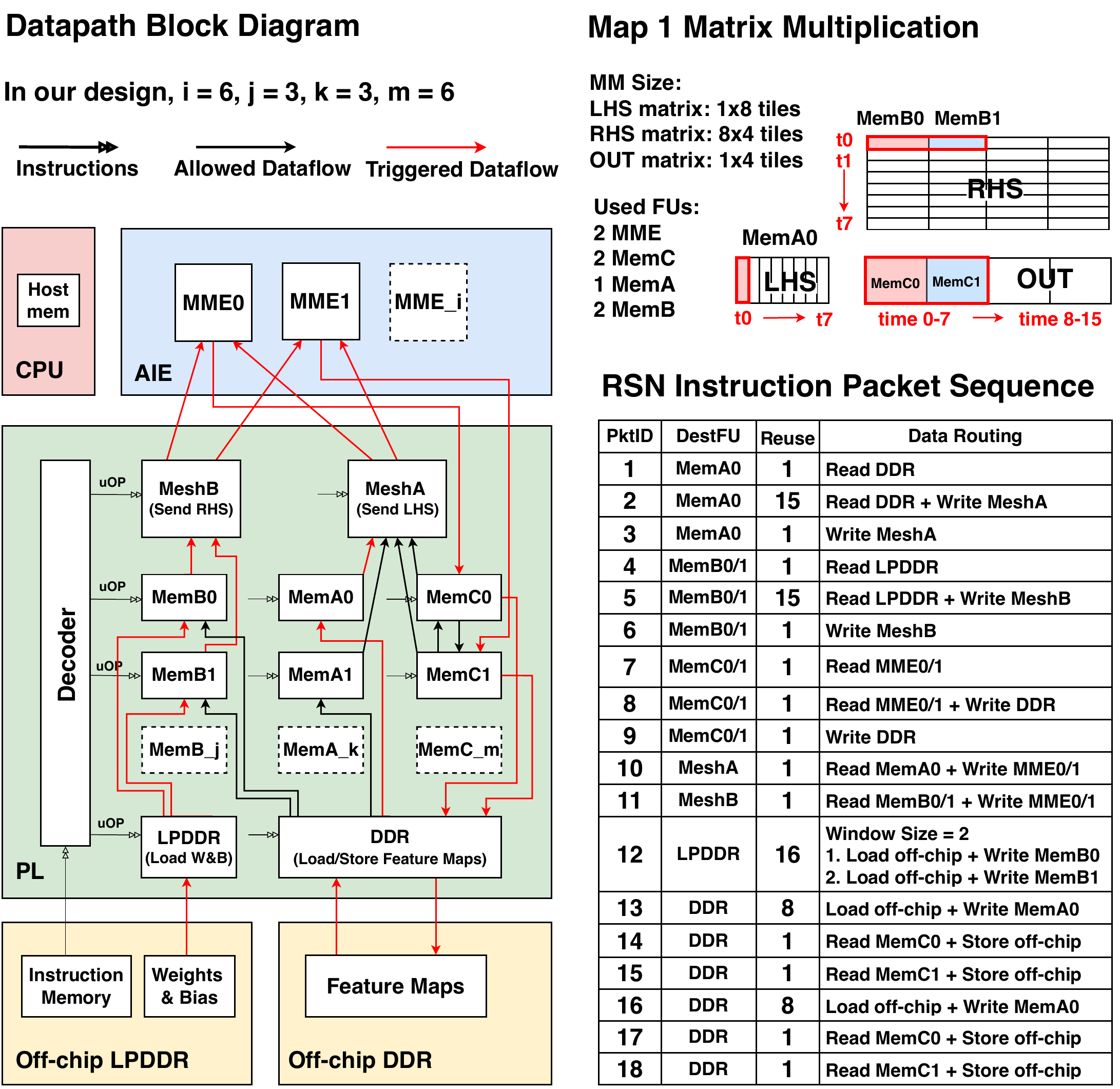}
    \vspace{-15pt}
    \caption{RSN-XNN Datapath and Example Application}
    \label{fig:blocks}
    \vspace{-15pt}
\end{figure}

\section {RSN-XNN: A RSN Case Study}

\subsection {Datapath and Instruction Set Overview}

We implement a proof-of-concept RSN design for transformer encoders, called RSN-XNN. 
Fig.~\ref{fig:blocks} depicts its datapath.
The AIE side has six matrix multiplication engine (MME) FUs that receive streaming inputs for the LHS operands from MeshA FUs and for the LHS operands from MeshB FUs, and send the results to MemC FUs on the PL side. Additionally, LPDDR FU loads weights and bias from off-chip LPDDR, while DDR FU manages the loading and storing of feature maps from off-chip DDR. 
The decoder unit fetches the RSN instruction sequence and issues uOP sequences to the FUs on the PL side. Since AIE tiles are processors with their own instruction memory, the uOPs for MME FUs are pre-stored locally and are not interleaved into the main single instruction sequence.

Table \ref{table:ISA} lists the control planes for different types of FUs in RSN-XNN. 
MME FUs mainly perform MMs in a tiled manner and also support several non-MM operations, including adding bias, adding the output of a previous layer to the current layer, and applying scale and shift in LayerNorm.
DDR/LPDDR FUs route data between off-chip and on-chip memory and MeshA/B FUs route data between AIE and on-chip memory. 
A fine-grained load/store interleaving can be achieved by orchestrating the uOP sequence for DDR. 
MME FUs can be flexibly grouped by modifying data routing in MeshA/B FUs, enabling the use of all six MME FUs for a single MM or the pipelined execution of multiple MMs.
MemA/B/C serves as flexible and fast scratchpads to increase on-chip data reuse, and they are double buffered to allow the overlapping of computation and data movement. They also implement some non-MM operations, such as Softmax, GELU, transpose, and the mean, variance, and normalization operations in LayerNorm.

Fig.~\ref{fig:blocks} also presents an example application performing a 1x8x4 tile MM, which triggers 2 MME, 1 MemA, 2 MemB, and 2 MemC FUs. 
MemA0 first expects the arrival of the first tile of LHS data from DDR FU. 
Then, it sends the previous tile to MeshA and receives the new tile from DDR, repeating this process 15 times. 
Finally, it sends the last tile.
MemB0/1 and MemC0/1 FUs are controlled similarly, using three instructions to manage the prolog, steady state, and epilog phases, respectively.
MeshA/B FUs fan in and fan out data between MemA/B and MME FUs.
MeshA reads LHS data from MemA0 and copies it to both MME0/1, while MeshB passes RHS data from MemB0 to MME0 and from MemB1 to MME1. Their actions are only set once because the dataflow remains the same. 
LPDDR FU loads 16 tiles of RHS data and alternates sending them to MemB0/1 FUs.  
DDR FU operates the same as Way 1 in Fig.~\ref{fig:interleaveRW}, storing 2 OUT tiles per 8 LHS input tiles loaded.

\begin{table}[t] 
\caption{uOP Control Planes Managing FUs in RSN-XNN}
\centering
\resizebox{\linewidth}{!}{
\begin{tabular}{p{1cm}p{8.5cm}}
\toprule
\textbf{FU} & \textbf{Control Plane}  \\
\midrule
MME & matrix size, tile size, add bias \ding{51}/\ding{55}, add previous layer \ding{51}/\ding{55}, calculate scale and shift \ding{51}/\ding{55}, accumulate along k \ding{51}/\ding{55}.	\\
\hline
DDR & addr, stride size, stride offset, stride count, load \ding{51}/\ding{55}, destFU, store \ding{51}/\ding{55}, srcFU.	\\
\hline
LPDDR & addr, stride size, stride offset, stride count, destFU, load bias \ding{51}/\ding{55}.	\\
\hline
MeshA/B & size, srcFUs, destFUs.	\\
\hline
MemA & matrix size, tile size, srcFU, load data \ding{51}/\ding{55}, send to MME \ding{51}/\ding{55}.	\\
\hline
MemB & matrix size, tile size, load data \ding{51}/\ding{55}, send to MME \ding{51}/\ding{55}, transpose input \ding{51}/\ding{55}, load bias \ding{51}/\ding{55}.	\\
\hline
MemC & matrix size from MME, matrix size to DDR, tile size from MME, tile size to DDR, receive from MME \ding{51}/\ding{55}, send to MME \ding{51}/\ding{55}, softmax \ding{51}/\ding{55}, gelu \ding{51}/\ding{55}, mean/variance/normalization \ding{51}/\ding{55}.\\
\bottomrule
\end{tabular}
}
\vspace{-10pt}
\label{table:ISA}
\end{table}

\subsection {Decision Process of Datapath Generation}
\label{section:decision process}
%


The datapath generation process includes three main stages:

\textbf{Model segmentation:}
We begin with a first-order formula-based calculation to segment targeted models to map resources efficiently. 
Compute-bound layers are segmented individually, whereas multiple memory-bound layers are grouped together and executed in a pipelined manner to reduce on-chip data accesses, as detailed in Section \ref{section:computemap}.
Moreover, multiple layers can be grouped to overlap the prolog and epilog phases of layers, as detailed in Section \ref{section:bandwidthmap}.

\textbf{Single model segment analysis:}
This stage involves analyzing data movement and transformations segment by segment.
For each segment, we decide on the on-chip buffer size for each operand to ensure sufficient reuse of on-chip data, computation resource allocation across different layers, datapath to fuse matrix multiplication and non-MM operations, and 
data layout transformations between PL/off-chip and PL/AIE.

\textbf{Collective datapath construction:}
This stage reviews all segments collectively to determine a “union” datapath that encompasses all segment requirements while minimizing unnecessary edges and FUs.
We initially assume a fixed-function style for the entire datapath, aiming to split the datapath and create new FUs only when divergent datapath reuse patterns are necessary.
For example, Mesh FUs are not created for outputs returning from AIE to PL, as each MME consistently communicates with the same MemC.
Also, although each MemA FU has local memories that can process 256 floats in parallel, we opt not to further partition it because no segment demands more fine-grained data movement patterns.
Furthermore, we expose only the necessary controls for datapath reuse to the ISA.
For instance, we do not expose data layout transformations from the PL side to the AIE side because the layout transformations remain the same across all segments.

\begin{table}[t]
\centering
\caption{Latency Estimation for Four Mapping Types}

\begin{tabular}{ p{0.3cm}p{0.5cm}p{1cm}p{0.5cm}p{1.5cm}p{1.07cm}p{0.8cm}}
\toprule
& & \scriptsize \textbf{Latency if \newline inf. FLOPS} 
& \scriptsize \textbf{Used AIE} 
& \scriptsize \textbf{Latency per MM \newline if inf. BW} 
& \scriptsize \textbf{Latency if \newline inf. BW} 
& \scriptsize \textbf{Final \newline Latency} \\ \midrule
\textbf{A} & MM1 & \multirow{2}{1.2cm}{2.22 ms} & \multirow{2}{0.3cm}{64\%} & 0.81 ms & \multirow{2}{1.2cm}{2.43 ms} & \multirow{2}{1.2cm}{2.43 ms} \\ \cline{2-2} \cline{5-5}
 & MM2 &  &  & 1.62 ms &  &  \\ \hline
\textbf{B} & MM1 & \multirow{2}{1.2cm}{10.9 ms} & \multirow{2}{0.3cm}{64\%} & 0.81 ms & \multirow{2}{1.2cm}{2.43 ms} & \multirow{2}{1.2cm}{10.9 ms} \\ \cline{2-2} \cline{5-5}
 & MM2 &  &  & 1.62 ms &  &  \\ \hline
\textbf{C} & MM1 & \multirow{2}{1.2cm}{10.9 ms} & \multirow{2}{0.3cm}{96\%} & 0.54 ms & \multirow{2}{1.2cm}{1.08 ms} & \multirow{2}{1.2cm}{10.9 ms} \\ \cline{2-2} \cline{5-5}
 & MM2 &  &  & 0.54 ms &  &  \\ \hline
\textbf{D} & MM1 & \multirow{2}{1.2cm}{2.24 ms} & \multirow{2}{0.3cm}{96\%} & 0.81 ms & \multirow{2}{1.2cm}{1.62 ms} & \multirow{2}{1.2cm}{2.24 ms} \\ \cline{2-2} \cline{5-5}
 & MM2 &  &  & 1.62 ms &  &  \\ \bottomrule
\multicolumn{7}{p{8cm}}{
\small
Attention layer in BERT-Large, B=6, SeqLen=512. Softmax is ignored.} \\
\multicolumn{7}{p{7cm}}{
\small
MM1: Key\(\times \)Query, 512\(\times\)64\(\times\)512, Number=96.
} \\ 
\multicolumn{7}{p{7cm}}{\small
MM2: MM1's Output\(\times\)Value, 512\(\times\)512\(\times\)64, Number=96.
} \\ 

\end{tabular}
\vspace{-10pt}
\label{tab:roofline-latency}
\end{table}

\subsection {Compute Resources Mapping}
\label{section:computemap}

Table~\ref{tab:roofline-latency} compares the performance of BERT's attention layer with different mapping types in Fig.~\ref{fig:map4} under the VCK190 hardware budget, estimated using the roofline model formula. It contains 96 independent attention heads, each with two dependent matrix multiplications, corresponding to the task and layer in Fig.~\ref{fig:map4}.
Types B and C have high latency due to large off-chip feature map accesses between the two MM layers.  
Types A and B both have low AIE utilization because the MMs are too small.  
Type D achieves the lowest latency with removed off-chip feature map accesses and high AIE utilization, while the extra latency caused by pipeline setup is negligible.  
Although this table shows the benefits of using the pipeline mapping type for attention layers, larger layers, such as the feedforward layers with large MMs, require another mapping style.  
Storing intermediate feature maps between two MMs in BERT-Large's feedforward layers requires over 25 MB of buffers, exceeding our on-chip capacity.  
However, these MMs are large enough to reach high AIE utilization when executing one MM at a time.
It can be seen that statically mapping two pipelined layers or mapping one layer at a time does not suit all cases.
It is crucial that RSN-XNN supports the flexibility for dynamically switching between mapping types at runtime, namely the dynamic chain of the pipelined FU in Table~\ref{tab:hardware_capabilities}.

\begin{figure}[t]
    \centering
\includegraphics[width=1.0\linewidth]{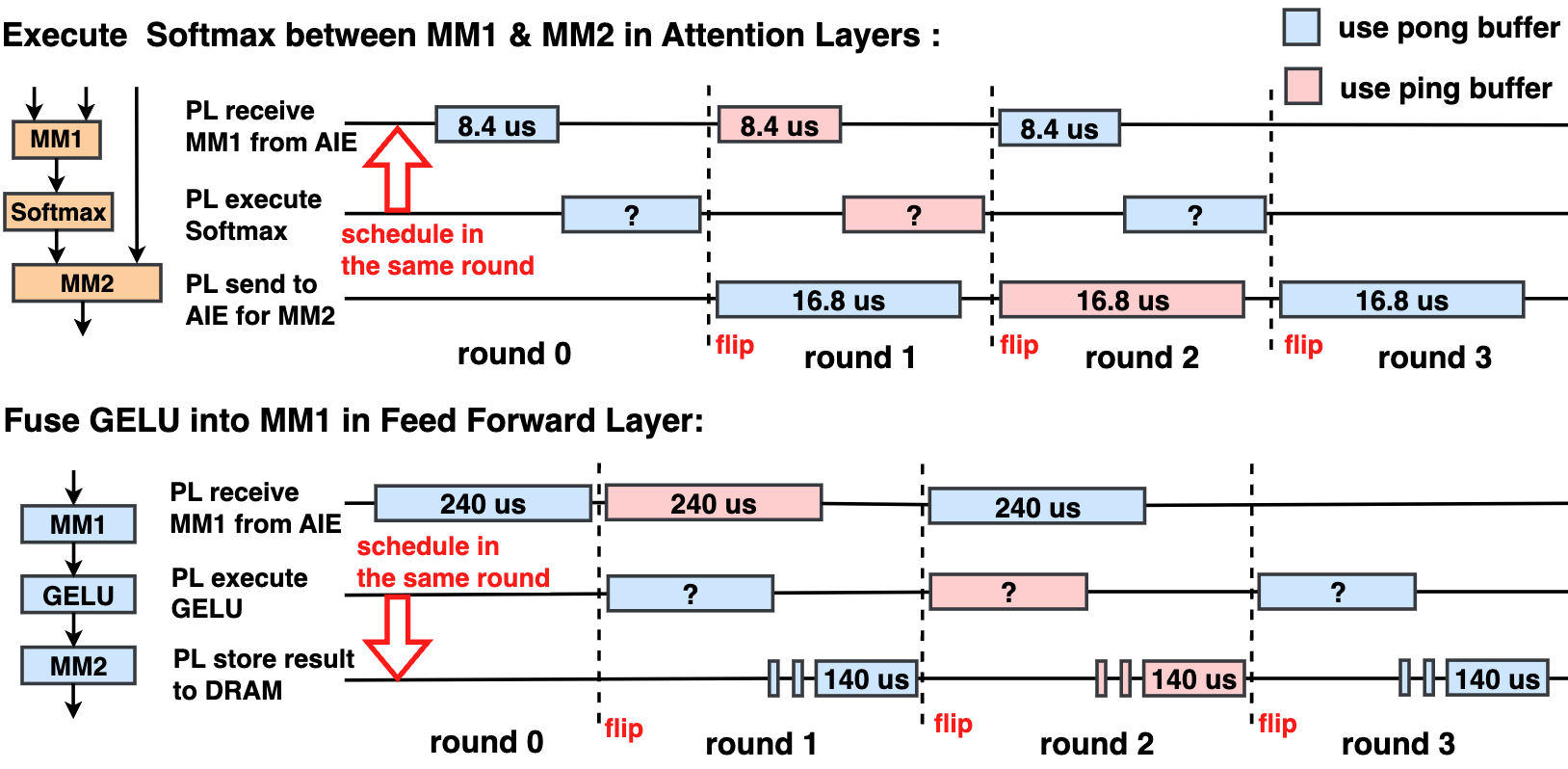}
\caption{Pipeline Non-MMs and Their Adjacent MMs }
    \label{fig:nonlinear}
    \vspace{-7pt}
\end{figure}

\begin{figure}[t]
    \centering
    \includegraphics[width=\linewidth]{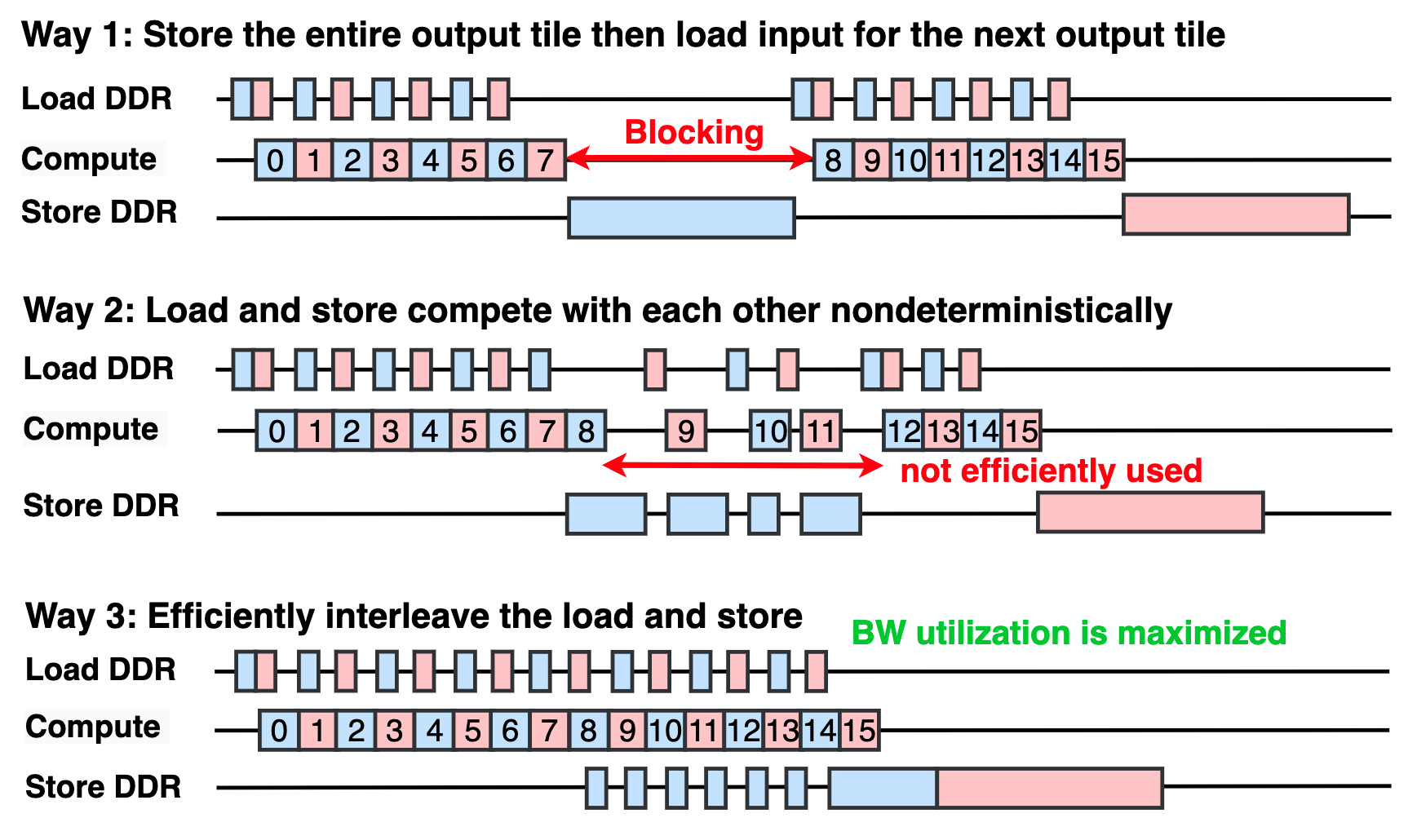}
    \caption{Three Ways to Map Load and Store Operations to One DDR Channel}
    \label{fig:interleaveRW}
    \vspace{-10pt}
\end{figure}

Non-MM operators are fused with their adjacent MM operations using a pipeline mapping style to avoid off-chip memory accesses and hide the latency of executing non-MMs within the time taken for MMs.  
As shown in Fig.~\ref{fig:nonlinear}, Softmax in attention layers occurs between two PL modules that:  
1) receive the 1st MM results from the AIE and 2) send the 2nd MM inputs to the AIE.  
A ping-pong buffer mechanism is used to allow overlapping between RCEV and SEND operations, enabling pipeline execution between the two MMs.  
We insert Softmax after RCEV and before the ping-pong buffers flip.  
This is because RCEV is shorter than SEND (8.4 vs. 16.8 µs for BERT-Large) in our mapping strategy, and scheduling Softmax together with RCEV utilizes the idle time during which RCEV waits for SEND to complete.  
The final throughput for processing the attention layers is determined by the maximum of the latency taken by RCEV plus Softmax, and the latency taken by SEND.  
Using a similar analysis, we insert the GELU operation right after the ping-pong buffers flip and before the start of storing the final result back to off-chip memory.


\subsection {Bandwidth Resources Mapping}

\label{section:bandwidthmap}

In RSN-XNN, instructions can explicitly specify \textbf{fine-grained off-chip load and store interleaving} to maximize the utilization of off-chip bandwidth.  
Fig.~\ref{fig:interleaveRW} shows an MM example where the input tiles are loaded from one DDR channel along the K dimension 8 times, followed by sending the completed output tile off-chip.  
If the load-compute-store order is strictly followed, the computation for the second output tile will stall when the first one is stored back to DDR.  
One way to improve this is to schedule the loading of 8 input tiles for the second output simultaneously with the storing of the first output tile.  
This can be achieved by pushing requests to the AXI read/write channels and letting the hardware controller arbitrate the execution.  
However, since the hardware controller does not have application-level information, it schedules loads and stores non-deterministically and cannot guarantee the optimal ordering.  
Instead, we explicitly specify the DDR load and store ordering using instructions, carefully interleaving the load/store operations to maximize bandwidth utilization.  
In addition to controlling execution within a single layer, we also support \textbf{overlapping the prolog and epilog phases of layers} by interleaving the storing of the last output tile from the previous layer with the loading of the first input tile for the current layer.  
This is particularly useful in scenarios involving many small layers, where the prolog and epilog phases are significant due to the small steady phases, such as in attention layers.  
Given that DNN inference is generally deterministic to allow for predictable memory utilization, it is beneficial to provide the flexibility for precise control of off-chip memory accesses. This enables optimal scheduling based on specific workloads and hardware capabilities.

\subsection {Compatibility of High-level Program}


\begin{figure}[t]
    \centering
    \includegraphics[width=\linewidth]{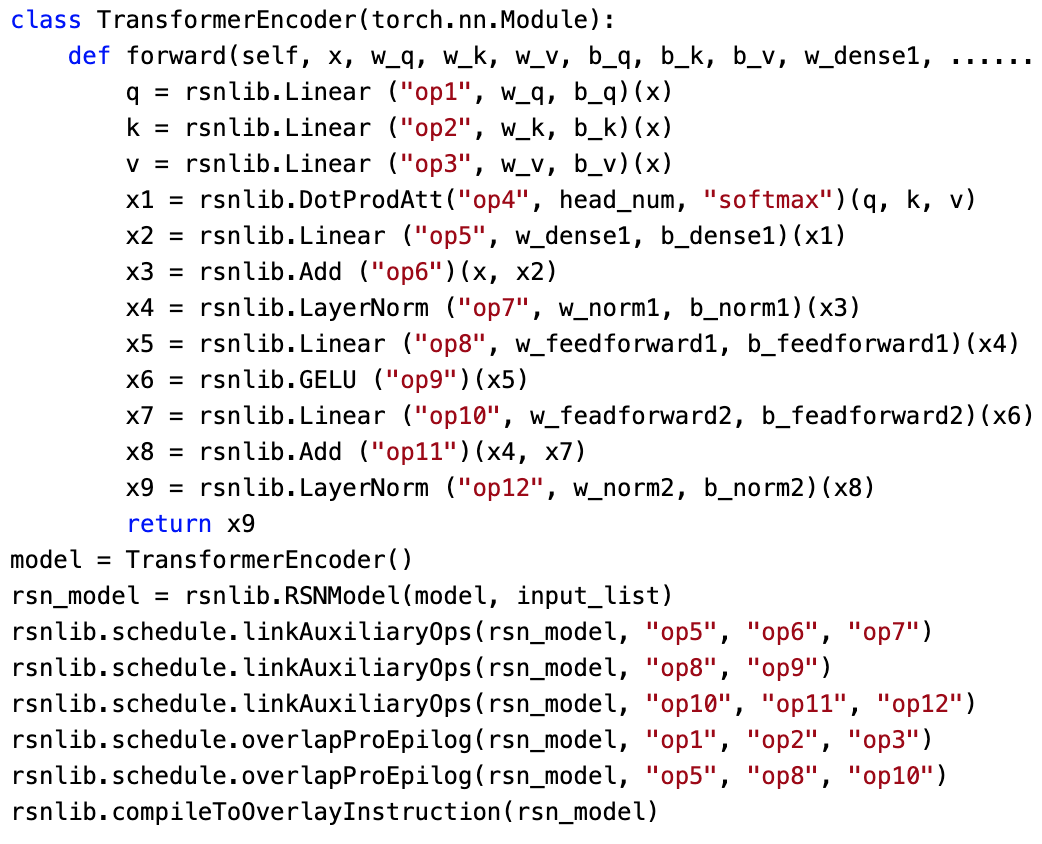}
    \caption{Domain Specific Library Usage Example}
    \label{fig:dsl}
    \vspace{-15pt}
\end{figure}

A domain-specific library, RSNlib, has been developed to generate RSN instructions from high-level Python code. 
Fig. \ref{fig:dsl} presents an example code that specifies the transformer encoder model. 
The library processes PyTorch models composed of RSNlib operators according to a predefined execution schedule. 
It employs a template-based approach to validate whether the model and schedule align with supported backend patterns. 
It shows that the proposed overlay software can be compatible with existing compiler infrastructures through library-based methods. 
Exploring the automatic generation of the datapath from arbitrary input code is beyond the scope of this paper but could be a topic for future research.

\section{Evaluation}

\textbf{Experiment setup:}
We implement RSN-XNN on the VCK190 Evaluation Kit~\cite{vck190} and program the PL side using HLS with the Vitis 2024.1 software~\cite{vitis2023}. 
We measure the execution latency on the CPU host using the std::chrono clock. 
All experiments use the same bitstream, varying the instructions passed to the datapath to accommodate different applications. 
We source inputs and weights for BERT-Large from Hugging Face~\cite{huggingface2023}, load them onto the board, and validate the outputs against reference results. 
We obtain latency for the T4, V100, and A100 GPUs from NVIDIA's state-of-the-art reports~\cite{nvidia_bert} and measure L4 GPU latency on Google Colab. 
We measure the GPU power using Pytorch and nvidia-smi and the VCK190 power using Xilinx's BEAM~\cite{xilinx_beam}.
We profile DRAM accesses with Pytorch and NVIDIA's Nsight Compute \cite{nvidia-ncu} on Google Colab.

\textbf{Precision:}
Experiments use FP32 precision. Although FP16 is preferred, the VCK190 supports only FP32, INT16, and INT8. INT16 is uncommon, and INT8 causes significant accuracy drops in BERT-Large, as Intel studied \cite{int8-drop}.

\begin{figure}[t]
    \centering
    \includegraphics[width=1\linewidth]{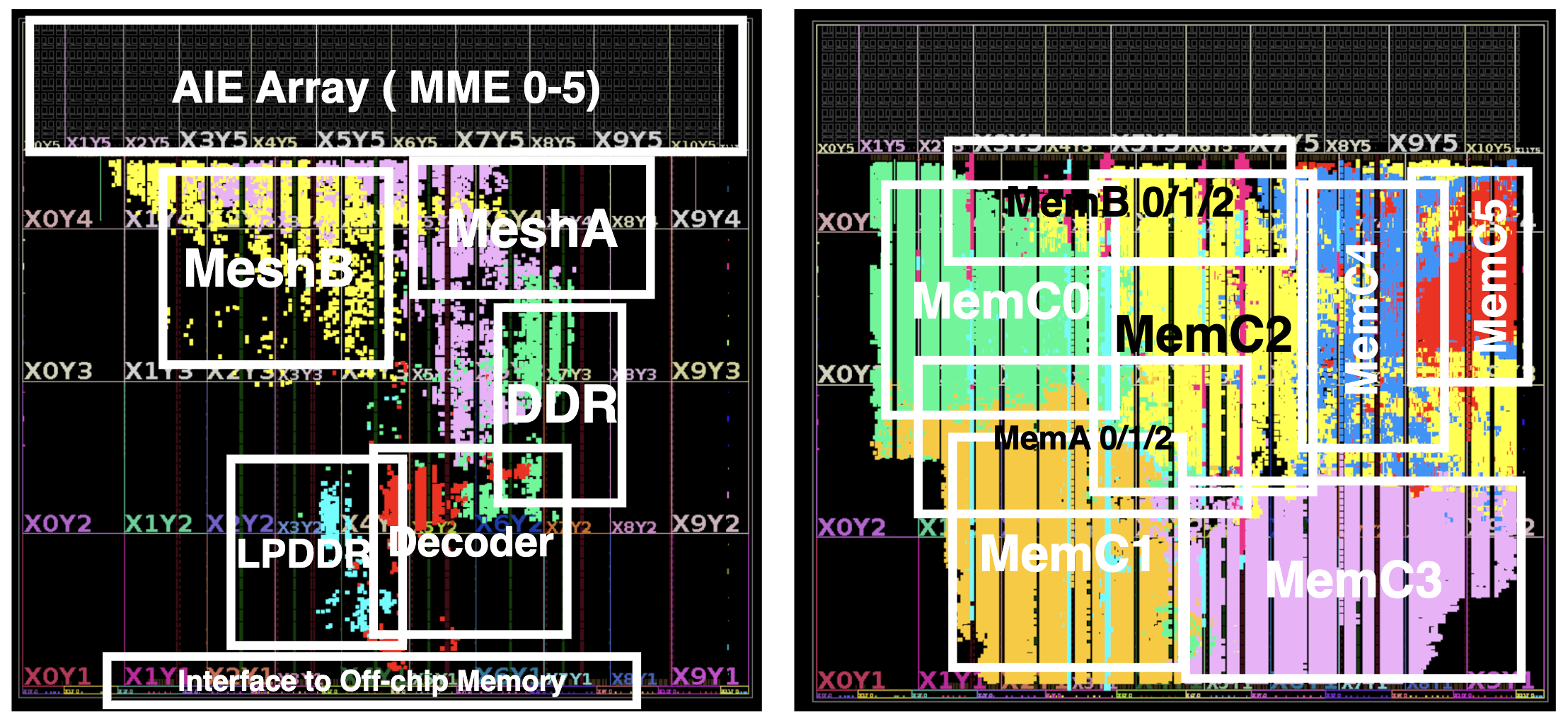}
    \caption{Device View of the Routed Design}
    \label{fig:impl}
    \vspace{-8pt}
\end{figure}

\begin{figure}[t]
    \centering
    \includegraphics[width=0.55\linewidth]{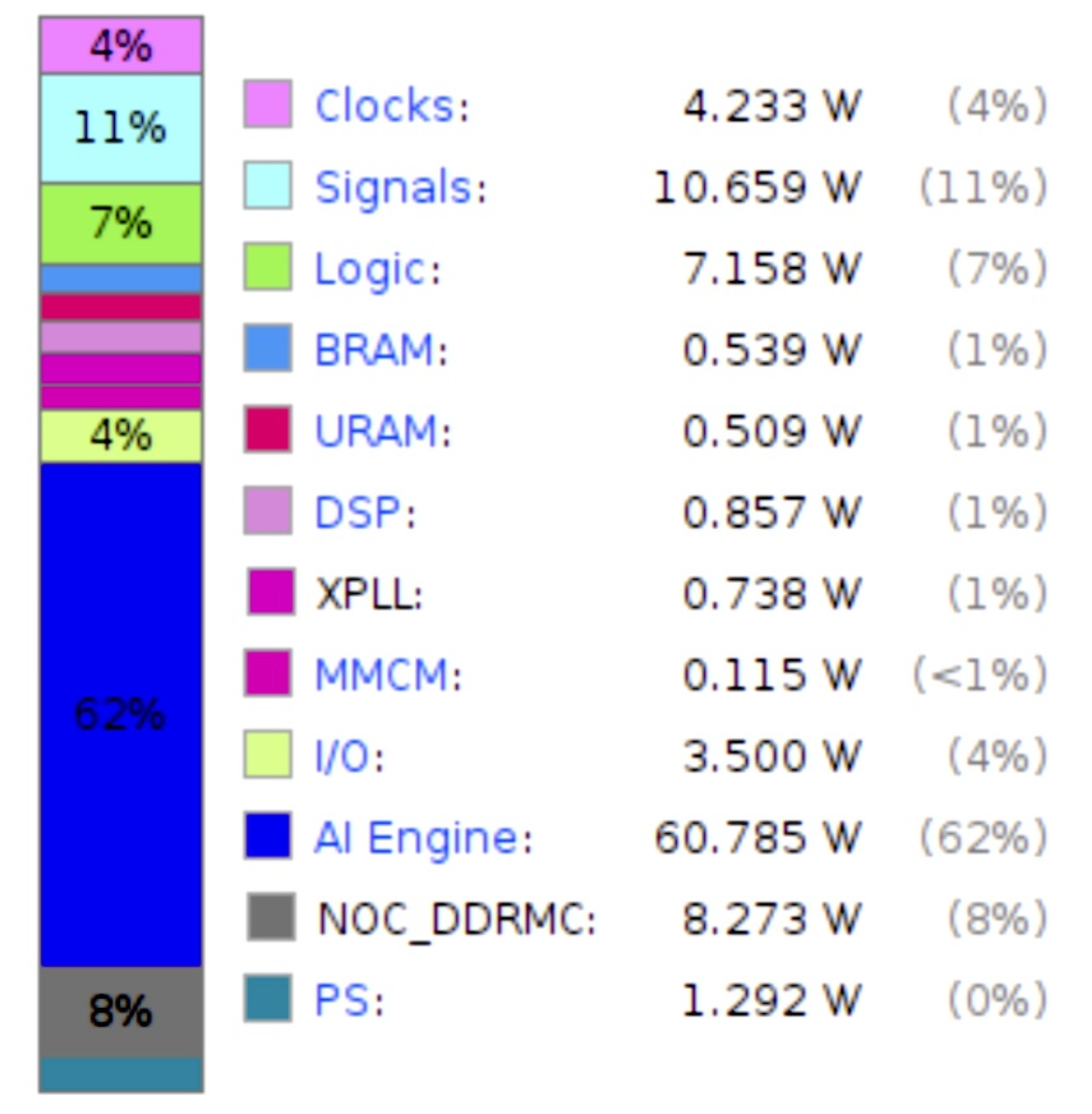}
    \caption{Power Estimation Summary (Total 98.66 W)}
    \label{fig:power}
    \vspace{-8pt}
\end{figure}

\textbf{Total area:}  
The FPGA runs at 260 MHz, and the AIE at 1.25 GHz.
Fig.~\ref{fig:impl} shows the routed design and Fig.~\ref{fig:power} shows the power estimation summary in Vivado. The final FPGA resource utilization includes 598,144 registers (33\%), 494,855 LUTs (55\%), 1,073 DSP blocks (55\%), 967 BRAMs (59\%), and 463 URAMs (41\%).


\subsection{Overhead Analysis}
\label{section:overhead}

We analyze the RSN ISA overheads from three perspectives.

\textbf{Energy:} 
Table \ref{tab:energybreakdown} presents the estimated power consumption for various components in the RSN-XNN, as reported by the Vivado power analysis \cite{vivado-power}.
These numbers are over-estimated in absolute terms, but provide valuable insights into the ratio of energy consumed by different components. On-board measurements cannot offer such a detailed breakdown.
AIE accounts for 62\% of the power. MemC FUs consume a large amount of power (23\%) due to the large on-chip memories and heavy arithmetic operations.
The energy overhead of the decoder unit is negligible (<0.08\%).

\begin{table}[t]
\centering
\caption{Estimated Power Consumption Breakdown for Decoder Unit and Different FU Types}
\vspace{-11pt}
\resizebox{\linewidth}{!}{
\label{tab:energybreakdown}
\begin{tabular}{p{0.4cm}p{0.9cm}p{0.4cm}p{0.66cm}p{0.66cm}p{0.66cm}p{0.4cm}p{0.8cm}p{0.7cm}p{0.7cm}}
\toprule
& \textbf{Decoder} & \textbf{AIE} & \textbf{MemC} & \textbf{MemB} & \textbf{MemA} & \textbf{DDR} & \textbf{LPDDR} & \textbf{MeshA} & \textbf{MeshB} \\
\midrule
\textbf{Watt} & 0.08 & 60.8 & 22.91 & 0.47 & 0.25 & 0.33 & 0.15 & 0.10 & 0.09 \\
\textbf{\%} & 0.08 & 61.6 & 23.22 & 0.48 & 0.25 & 0.33 & 0.15 & 0.10 & 0.09 \\
\bottomrule
\end{tabular}
}
\vspace{-5pt}
\end{table}

\begin{table}[t]
\caption{Overlay Area Overhead and Utilization Comparison }
\vspace{-8pt}
\label{table:overhead}
\resizebox{\linewidth}{!}{
\centering
\begin{tabular}{p{1.4cm}p{0.9cm}p{1.1cm}p{1.2cm}p{0.9cm}p{1cm}}
\multicolumn{6}{l}{\textbf{(a) Area Overhead of the Instruction Decoder (\% of Total Design)}}\\ \toprule
\textbf{Design} & \textbf{Device} & \textbf{LUT} & \textbf{FF} & \textbf{DSP} & \textbf{BRAM} \\
\hline
RSN-XNN   & VCK190 & 11.7k(3\%) & 8.6k(2.5\%) & 5(0.5\%)  & 4(0.3\%) \\ \hline
DFX       & U280   & 3k(0.6\%)   & 13k(1.2\%)  & 0         & 24(2\%) \\ \hline
DLA       & Arria10 & \multicolumn{4}{p{5cm}}{Use 2046 ALMs (7\% of total ALMs on board). Total design area is unreported. } \\ 
\bottomrule
\multicolumn{6}{p{8.4cm}}{ 
} \\ [-6pt]
\end{tabular}
}
\resizebox{\linewidth}{!}{
\begin{tabular}{p{1.4cm}p{0.75cm}p{1.5cm}p{1.5cm}p{2.05cm}p{0.5cm}}
\multicolumn{6}{l}{\textbf{(b) Computation Resource Utilization}}\\ \toprule
\textbf{Design} & \textbf{Precis.} & \textbf{Peak Perf. \newline (TFLOPS)} & \textbf{Off-chip \newline BW (GB/s)} & \textbf{Achieved Perf. \newline (TFLOPS)} & \textbf{Util.\newline(\%)} \\
\hline
RSN-XNN   & FP32 & 8   & 57.6 & 4.7   & 59 \\ \hline
DFX       & FP16 & 1.2 & 460 & 0.19 & 16 \\ \bottomrule
\multicolumn{6}{p{9.4cm}}{
DLA does not report achieved performance in FLOPS. \newline
Achieved throughput performances occur under similar transformer computations: RSN-XNN for BERT-Large and DFX for GPT-2 prefill stages.
} \\ 
\end{tabular}
}
\vspace{-15pt}
\end{table}

\textbf{Instruction:} 
Fig.~\ref{fig:instructionbyte} shows the size of the instructions and uOPs for BERT-Large. 
For this application, the PL side is programmed by a total of 1685 RSN instructions, distributed as follows: 1404 for DDR, 88 for LPDDR, 49 for MemAs, 58 for MemBs, 22 for MemCs, 38 for MeshA, and 26 for MeshB.
Each AIE tile relies on a 4-byte control input (uOP) to select its operations and is programmed with 17 uOPs in total, resulting in a negligible RSN instruction footprint to support its virtualization as part of an MME FU.
Additionally, the average RSN instruction processing rate of only 1.4 MB/s demonstrates that the overhead of using instructions on the PL side is quite small, merely 0.0024\% of the off-chip bandwidth.
The average compute-to-instruction ratio, accounting for both the PL and AIE sides, reaches up to 1.6 GFLOPs per byte, demonstrating that those instructions are highly efficient and the system requires minimal instruction-level intervention.

\textbf{Area:}  
Table \ref{table:overhead} presents the area overhead of instruction decoder units across different designs. 
The data shows that RSN-XNN maintains a low area overhead, both in absolute terms and as a percentage of the total design area. 
The table also includes data for two overlays: DLA\cite{overlay-dla}, which controls execution at the tile level, and DFX\cite{overlay-dfx}, which controls execution at the layer level.
RSN-XNN's area overhead is comparable to existing overlays but offers greater flexibility and better resource utilization.
As the instruction processing rate is very low, area can be saved by slowing down the decoder unit.
We employed several low-cost techniques, such as increasing the loop initiation interval and allowing multiple cycles to decode an instruction.

\subsection{Characterization of FU Properties}

\begin{figure}
    \centering
    \includegraphics[width=\linewidth]{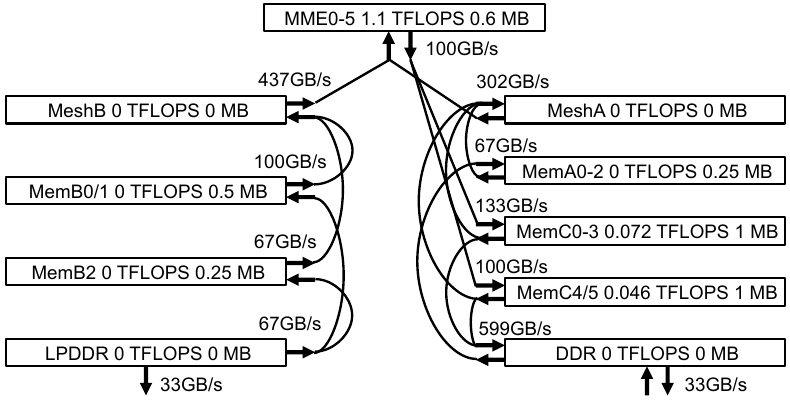}
    \caption{Visual Layout of FU Compute, Memory, and Bandwidth Properties}
    \label{fig:FUproperties}
    \vspace{-8pt}
\end{figure}

Fig.~\ref{fig:FUproperties} presents the properties of each FU in terms of memory capacity, maximum floating-point compute throughput, and aggregate communication bandwidth (i.e., the sum of its incoming and outgoing network edges). 
This visualization clearly reveals the wide heterogeneity and coarse granularity of RSN-XNN’s FUs. 
For example, the six MME FUs each provide 590 KB of on-chip storage and 1.1 TFLOPS of compute throughput, while MeshA and MeshB serve purely as communication routers without memory or computation. 
The hardware properties of each FU are explicitly exposed in the ISA abstraction, allowing software to easily exploit them for scheduling and optimization.

\subsection {GEMM Performance}

\begin{table}[t]
\centering
\caption{Matrix Multiplication Throughput Comparison}
\label{table:GEMM}
\vspace{-8pt}
\resizebox{\linewidth}{!}{
\begin{tabular}{p{1.9cm}p{1.7cm}p{1.25cm}p{1.6cm}p{1cm}}
\multicolumn{5}{l}{\textbf{(a) AIE MM Throughput (PL Generates Data Without DRAM)}}\\ \toprule
\textbf{Method} & \textbf{AIE TileSize \newline (MxKxN)} & \textbf{Used AIE} & \textbf{Throughput (GFLOPS)} & \textbf{Gain} \\ \hline
CHARM \cite{charm} & 32x32x32 & 384 (96\%) & 4504.46 & + 0\%\\ \hline
MaxEVA \cite{maxeva}& 32x32x32 & 390 (98\%) & 5442.11 & + 20.8\% \\ \hline
AMA \cite{ama}& 32x32x32 & 342 (86\%) & 5867.29 & + 30.3\% \\ \hline
RSN-XNN & 32x16x32 & 384 (96\%) & 6095.64 & \textbf{+ 35.3\%}\\ \hline
RSN-XNN & 32x32x16 & 384 (96\%) & 6306.02 & \textbf{+ 40.0\%}\\ \hline
RSN-XNN & 32x32x32 & 384 (96\%) & 6784.96 & \textbf{+ 50.6\%}\\ \bottomrule
\multicolumn{5}{p{8.4cm}}{ 
} \\ [-6pt]
\end{tabular}
}
\resizebox{\linewidth}{!}{
\begin{tabular}{p{2.5cm}p{2cm}p{1.6cm}p{1.2cm}}
\multicolumn{4}{l}{\textbf{(b) End-to-end MM Throughput (With DRAM)}}\\ \toprule
\textbf{Square MM Size} & \textbf{CHARM \cite{charm}} & \textbf{RSN-XNN} & \textbf{Gain} \\ \hline
1024 & 1103.46 & 2982.62 & + 170.3\%\\ \hline
3072 & 2850.13 & 6600.12 & + 131.6\%\\ \hline
6144 & 3277.99 & 6750.93 & + 105.9\%\\ \bottomrule
\multicolumn{4}{p{9cm}}{
CHARM only uses DDR memory. \newline
MaxEVA and AMA do not explore end-to-end throughput with DRAM.
} \\ 

\end{tabular}
}
\vspace{-5pt}
\end{table}

\begin{figure}[bt]
    \centering
    \includegraphics[width=1\linewidth]{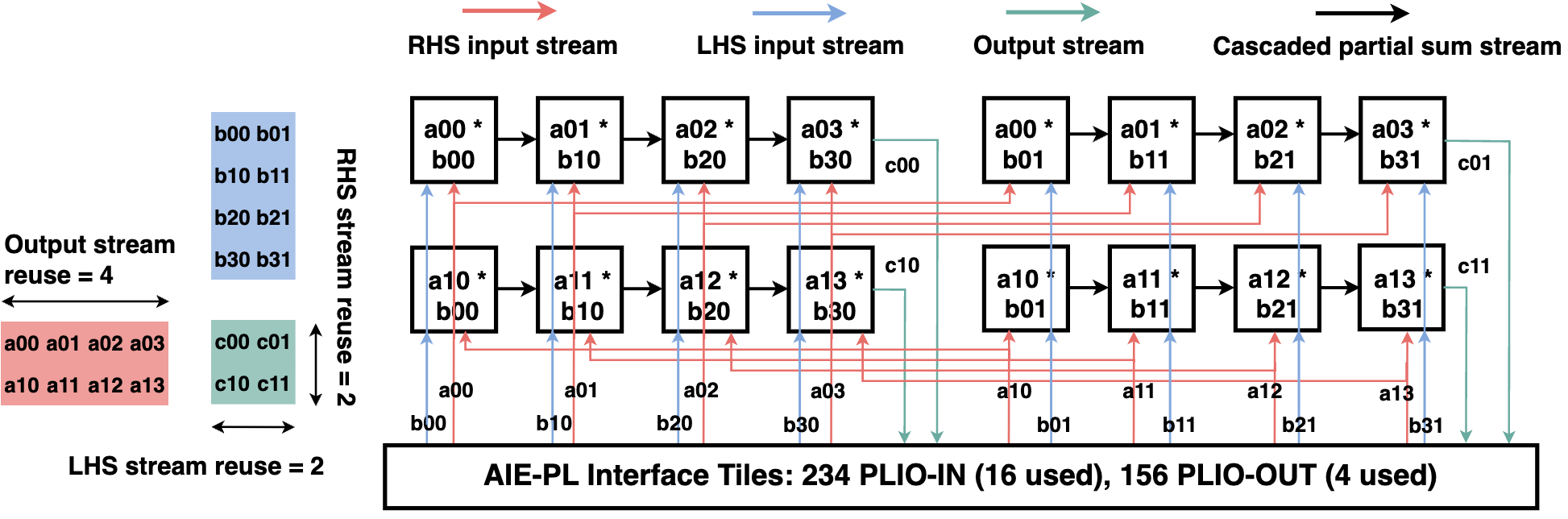}
    \caption{Reuse of AIE to/from PL Streams}
    \label{fig:aiearray}
    \vspace{-5pt}
\end{figure}

Table \ref{table:GEMM}
shows the performance of a single GEMM kernel with data either generated from the PL side or transferred from DRAM. Our AIE programming achieves 
a 16\% throughput improvement over the state-of-the-art AMA~\cite{ama}. 
Fig.~\ref{fig:aiearray} shows our optimization strategy. 
Each AIE tile uses two input streams and one output stream.  
With 400 AIE tiles in VCK190, using all of the tiles requires 800 input streams and 400 output streams, but VCK190 allows for only 234/156 64-bit input/output streams between the AIE and PL.
To address this, we group AIE tiles to share input streams. Output streams are reduced by chaining tiles together, cascading data to the next tile instead of sending it to the PL.  
We create 6 groups, each corresponding to one MME FU and containing 64 tiles in a 4x4x4 format, reusing LHS/RHS/output streams 4 times.
This setup utilizes 384 AIE tiles, 192 input, and 96 output streams, all within the resource budgets.

Although off-chip memories theoretically offer 25.6 GB/s for DDR and 32 GB/s for LPDDR, the peak observed bandwidths are 21 GB/s (DDR reads), 23.5 GB/s (DDR writes), and 20.5 GB/s (LPDDR reads).  
To reach the peak performance of 6.78 TOPS, each loaded weight must be reused over 661 times.  
DDR is used for both feature map loading and storing, requiring even larger reuse.  
We implement an output-stationary MM tiling scheme on the PL side, with the LHS tile size set to 768x128, the RHS to 128x1024, and the output to 768x1024, allowing for complete accumulation along the K dimension before storing off-chip.  
This setup enables 768x reuse of RHS, 1024x reuse of LHS, and an efficient output accumulation. 
We optimize further by finely interleaving load and store operations using RSN instructions.
To reduce strided off-chip memory accesses, data is stored in a 128x64 blocked layout off-chip, and MemA/B/C handle on-chip conversion from blocked to row-major or transposed format.  
Our solution is more effective than ~\cite{charm}.

\begin{figure}[t]
    \centering
    \includegraphics[width=1\linewidth]{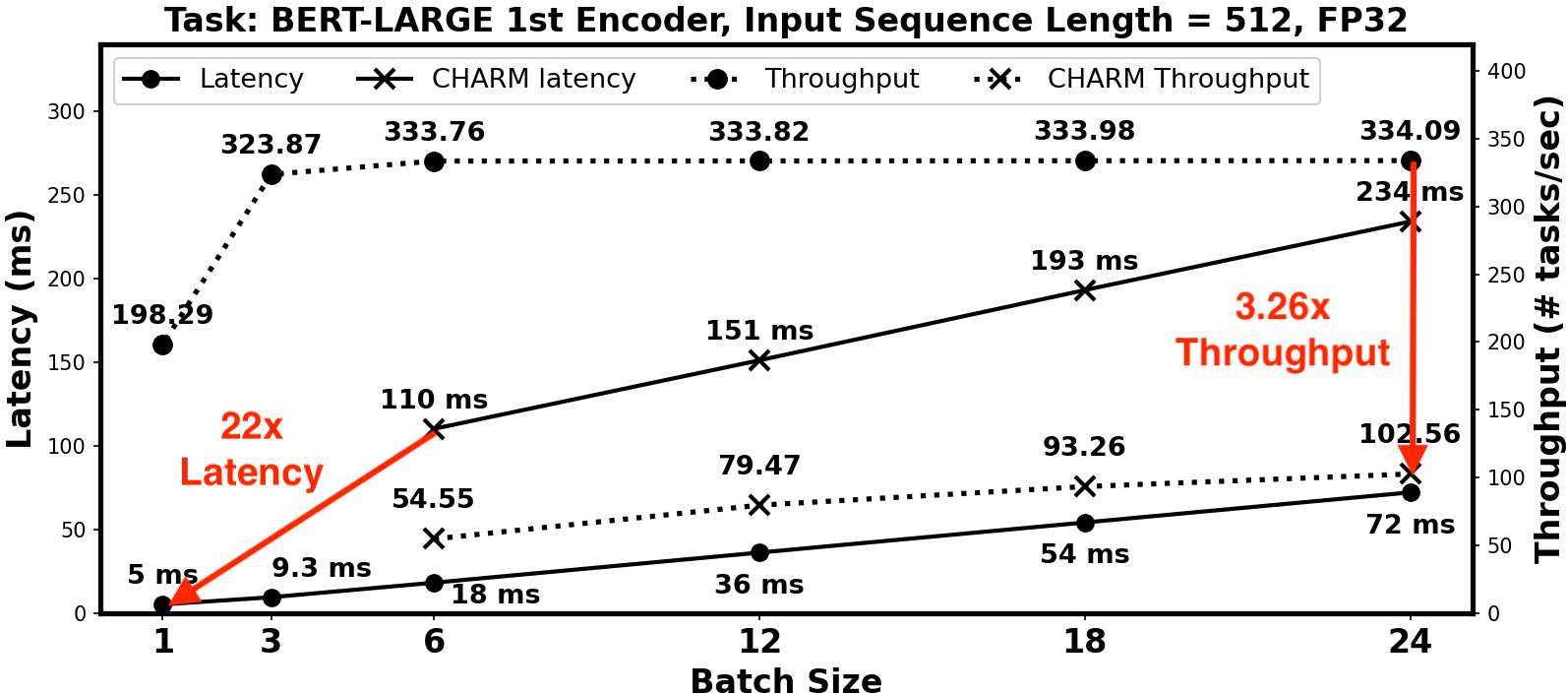}
    \vspace{-15pt}
    \caption{Achieved Latency/Throughput VS CHARM~\cite{charm}}
    \label{fig:input512}
    \vspace{-5pt}
\end{figure}

\begin{table}[t]
\centering
\caption{Comparison of Latency per Task at Maximum Throughput}
\vspace{-5pt}
\label{tab:latency_comparison}
\vspace{-5pt}
\begin{tabular} { ccccc }
\toprule
\textbf{Model} & \textbf{BERT} & \textbf{VIT} & \textbf{NCF} & \textbf{MLP} \\ \midrule
CHARM          & 57.2 ms       & 57.7 ms      & 40.4 ms      & 119 ms       \\ \hline
RSD-XNN        & 17.98 ms      & 23.7 ms      & 16.1 ms      & 42.6 ms      \\ \bottomrule
\end{tabular}
\vspace{-5pt}
\end{table}

\begin{table}[t]
\centering
\caption{Comparison of Maximum Throughput of Different SOTA FPGA-Based Transformer Accelerators} 
\resizebox{1\linewidth}{!}{
\label{tab:sota-fpgas}
\begin{tabular}{p{1.8cm}p{1cm}p{0.6cm}p{0.6cm}p{0.6cm}p{0.5cm}p{1.6cm}} 
\toprule
 \textbf{Design} & \textbf{Board} & \textbf{Prec.}  & \textbf{Peak TOPS} & \textbf{Achi. TOPS} & \textbf{Util. (\%)} & \textbf{Model} \\
\midrule
RSN-XNN & VCK190 & FP32 & 8  & 4.7 & 59 & BERT-L \\
SSR \cite{SSR_2024} & VCK190 & INT8 & 102 & 26.7 & 26 & DeiT-T \\
FET-OPU \cite{fet-opu-u280} & U280 & INT8 & 7.2  & 1.64 & 23 & BERT-B \\
DFX \cite{overlay-dfx}& U280 &  FP16 & 1.2  & 0.19 & 15 & GPT2 Prefill \\
VIA \cite{via-u50}& U50 & FP16 & 1.2  & 0.31 & 26 & Swin-T \\
FTRANS \cite{FTRANS} & VCU118 & INT16 & 2.7 & 1.05 & 38 & RoBERTa-B \\
\toprule
\multicolumn{7}{p{9cm}}{
\small{
U280, U50, and VCU118 do not have AIE. 
VIA operates at 300MHz; CAT, FET-OPU, and DFX at 200MHz. 
We assume FTRANS's frequency is 200MHz as it is unreported. 
In general, a higher achieved frequency on the same board suggests a more optimized design. 
Here the peak TOPS is calculated based on DSP count and the achieved frequency, as no standard best achievable frequency exists.
}
} \\ 

\end{tabular}
}
\vspace{-15pt}
\end{table}

\subsection {Comparison to SOTA FPGA Design}


\begin{table*}[t]
\caption{Execution Details of Different Model Segments}
\vspace{-5pt}
\footnotesize
\resizebox{\textwidth}{!}{
\begin{tabular}{p{1.6cm}p{2.4cm}p{2.6cm}p{2cm}p{2.2cm}p{2.8cm}p{1.3cm}}
\multicolumn{7}{c}{BERT-Large 1st Encoder, Sequence Length = 512, Batch = 6, FP32.
}\\\toprule
\textbf{MMs} & \textbf{Size (M\(\times\)K\(\times\)N\(\times\)Num)} & \textbf{Combined non-MMs} & \textbf{No Optimize (ms)} & \textbf{BW Optimized (ms)} & \textbf{Multi MMs together (ms)} & \textbf{Final (ms)} \\ \hline
Key & 3072\(\times\)1024\(\times\)1024\(\times\)1 & Bias & 1.667 (1x) & 1.276 (1.31x) & \multirow{3}{2.8cm}{3.584 (1.4x) \newline Overlap prolog/epilog}  & \multirow{10}{0.9cm}{17.98 \newline (2.47x)}  \\ \cline{1-5}
Query & 3072\(\times\)1024\(\times\)1024\(\times\)1 & Bias & 1.667 (1x) & 1.276 (1.31x) &  &  \\ \cline{1-5}
Value & 3072\(\times\)1024\(\times\)1024\(\times\)1 & Bias & 1.667 (1x) & 1.276 (1.31x) &  &  \\ \cline{1-6}
Attention MM1 & 512\(\times\)64\(\times\)512\(\times\)96 & Transpose, Softmax & 10.55 (1x) & -- & \multirow{2}{2.9cm}{ 2.618 (8.52x) \newline Pipeline MMs + overlap} &  \\ \cline{1-5}
Attention MM2 & 512\(\times\)512\(\times\)64\(\times\)96 &  & 11.75 (1x) & -- &  &  \\ \cline{1-6}
Dense & 3072\(\times\)1024\(\times\)1024\(\times\)1 & LayerAdd, Scale \& Shift, \newline Bias, Mean \& Var, Norm & 2.913 (1x) & 2.035 (1.43x) & \multirow{6}{2.8cm}{11.88 (1.45x) \newline Overlap prolog/epilog}  &  \\ \cline{1-5}
Feedforward MM1 & 3072\(\times\)1024\(\times\)4096\(\times\)1 & Bias, GELU & 8.492 (1x) & 5.501 (1.55x) &  &  \\ \cline{1-5}
Feedforward MM2 & 3072\(\times\)4096\(\times\)1024\(\times\)1 & LayerAdd, Scale \& Shift, \newline Bias, Mean \& Var, Norm & 5.764 (1x) & 4.811 (1.20x) &  &  \\ \bottomrule
\end{tabular}
\label{table:mm_operations}
}
\end{table*}

Fig.~\ref{fig:input512} compares the latency and throughput of RSN-XNN against the state of the art~\cite{charm} at varying batch sizes for the first encoder of BERT-Large.
Our best latency is 5 ms at batch size (B)=1, which is 22x faster than their best latency of 110 ms at B=6.  
At the same batch size, we achieve speedups ranging from 6.1x (B=6) to 3.3x (B=24).  
Regarding throughput, our performance nearly saturates at B=3 (97\% of the peak) and reaches a peak of 333.76 tasks/sec at B=6, which is 3.25x better than CHARM's best throughput at B=24.  
Their approach designs two separate MM engines for small and large layers, requiring them to schedule at a 6-batch granularity and interleave the execution of four 6-batches to fully overlap small and large MME executions.  
In contrast, RSN-XNN dynamically switches between pipeline and non-pipeline execution to efficiently handle small and large layers, allowing for single-batch execution.  
In addition, CHARM has to move the intermediates between the two MMs in the attention layers off-chip because it cannot support layer pipelining, which results in larger off-chip accesses.

Table \ref{tab:latency_comparison} compares the latency per task at maximum throughput for four applications: BERT, VIT, NCF, and MLP, against CHARM. 
All task size configurations align with CHARM’s implementations.
RSN-XNN achieves throughput improvements of 3.2x, 2.4x, 2.5x, and 2.8x for BERT, VIT, NCF, and MLP. 
Moreover, CHARM necessitates redesigning the datapath for different applications, whereas our design uses the same datapath for all applications.

Table \ref{tab:sota-fpgas} compares the FPGA-based transformer accelerators. Although our absolute performance is lower compared to SSR \cite{SSR_2024}, which also targets VCK190 but operates at INT8 precision, our utilization of peak performance is more efficient. For pure FPGA-based designs, our approach has much larger operations per second thanks to the very high peak performance provided by AIE.


\subsection {Comparison to Baseline Overlay Style}
\label{sec:latencybreakdown}

Table~\ref{table:mm_operations} provides a latency breakdown of different model segments and the effects of different optimization techniques. 
With fine-grained load and store interleaving, 
the first three large MMs achieve a 1.31x latency speedup, while the last three large MMs achieve speedups of 1.43x, 1.55x, and 1.20x, respectively.
Small MMs in the attention layers struggle with sequential execution due to 
large off-chip access for storing/loading feature maps, low MME utilization, and a short steady-state for overlapping memory access and computation.
We observed a 8.52x speedup by executing attention MM1 and MM2 in a pipelined manner and overlapping the prolog/epilog across parallel attention heads.
Compared to a typical overlay style that executes layers sequentially without fine-grained bandwidth mapping, RSN-XNN achieves a 2.47x speedup.

\subsection {Comparison to GPUs}

Table~\ref{table:comparison_hardware} compares the performance and energy efficiency of RSN-XNN with the NVIDIA T4~\cite{nvidia_t4}, V100~\cite{nvidia_v100}, A100~\cite{nvidia_a100}, and L4 GPUs~\cite{nvidia_l4}.   
On the VCK190, we execute the encoder layer 24 times to obtain latency, with the embedding layers considered negligible (less than 0.2 ms on 
the T4).
For the T4 GPU with the same 8 TFLOPS FP32 performance, RSN-XNN achieves slightly better latencies at B=2, B=4, and B=8, despite the VCK190 having only 58 GB/s of off-chip memory bandwidth compared to the T4’s 320 GB/s.
When B=1, our latency is worse than the T4’s (0.7x) because the small matrix size limits weight reuse to 384 times, only half of the 661 times needed for peak performance due to our low off-chip memory bandwidth. 
For the A100 GPU under the same 7nm process node, RSN-XNN achieves a 2.1x/4.5x better operating/dynamic energy efficiency in FP32 but has a slower latency because of A100's superior peak performance and bandwidth.  
Compared to the modern energy efficient L4 GPU, RSN-XNN is 0.69x slower but slightly more energy efficient in FP32. 
All GPUs should reach saturation in FP32 at B=8, as indicated by the latency trend with increasing batch size.
These statistics offer quantitative insights into the performance and energy differences between an efficient Versal design and GPUs for a large MM workload.
To furthur explain our energy efficiency advantages over NVIDIA's T4 and A100 GPUs, we profile total DRAM accesses. 
Compared to T4 and A100 GPUs, our RSN-XNN exhibits significantly fewer off-chip memory accesses (2.6/2.8x reduction), which contributes to our improved energy efficiency. 
RSN-XNN reuses data moved on-chip  very efficiently and supports pipelined execution to avoid moving intermediate data off-chip.
We also offer a data point using the A100 in FP16, showing much better latency and twice the energy efficiency due to its FP16 performance being 39 times greater than the VCK190's peak FP32 performance. This underscores the need for FPGAs to continue integrating ASIC efficiency for enhanced performance and bandwidth.

\begin{table}[t]
\centering
\caption{Comparison of T4, V100, A100, L4, and VCK190}
\vspace{-5pt}
\label{table:comparison_hardware}
\footnotesize
BERT-Large, Sequence Length = 384, FP32.
\resizebox{\linewidth}{!}{
\begin{tabular}{p{2.35cm}p{0.45cm}p{0.48cm}p{0.48cm}p{0.48cm}p{0.45cm}p{1.1cm}}
\toprule
\textbf{} & \textbf{T4} & \textbf{V100} & \textbf{A100} & \textbf{A100} & \textbf{L4} & \textbf{VCK190} \\ 
\midrule
\textbf{Precision} & FP32 & FP32 & FP32 &FP16 & FP32 & FP32 \\ \hline
\textbf{Release Date} & 2018 & 2017 & \multicolumn{2}{ p{0.96cm}}{ \centering 2020} & 2023 & 2021 \\ \hline
\textbf{Process (nm)} & 12 & 12 & \multicolumn{2}{p{0.96cm}}{\centering7} & 5 & 7 \\ \hline
\textbf{Peak Perf. (TFLOPS)} & 8.1 & 15.7 & 19.5 & 312 & 30.3 & 8.0 \\ \hline
\textbf{Off-chip BW (GB/s)} & 320 & 900 & \multicolumn{2}{p{0.96cm}}{\centering1555} & 300 & 57.6 \\ \hline
\textbf{Die Area (mm\textsuperscript{2})} & 545 & 815 & \multicolumn{2}{p{0.96cm}}{\centering826} & 294 & $\leq$ 458~\cite{amd_versal_die_area}\\ 
\hline
\multicolumn{7}{l}{\textbf{Latency (ms) by Batch Size}} \\  
\textbf{B = 8} & 499 & 182 & 137 & 23 & 307 & 444 \\  
\textbf{B = 4} & 258 & 93 & 72 & 15 & 156  & 220 \\  
\textbf{B = 2} & 127 & 49 & 40 & 10 & 83 & 122 \\  
\textbf{B = 1} & 67 & 29 & 23 & 8 & 41 & 95 \\ 
\hline
\multicolumn{7}{l}{\textbf{Energy Efficiency (Batch Size = 8)}} \\ 
\textbf{Operating Power (W)} & 72 & 292 & 308 & 392 & 72 & 45.5 \\ 
\textbf{Dynamic Power (W)} & 42 & 256 & 268 & 352 & 41 & 18.2\\ 
\textbf{Opt. Efficiency (Seq/J)} & 0.22 & 0.15 &  0.19 & 0.89 & 0.36 & 0.40 \\  
\textbf{Dy. Efficiency (Seq/J)} & 0.38 & 0.17 &  0.22 & 0.99 & 0.64 & 0.99\\ 
\hline
\multicolumn{7}{l}{\textbf{Off-Chip DRAM Usage (Batch Size = 8)}} \\ 
\textbf{Total Accesses (GB)} & 31 & - & 34 & 25 & 12 & 12 \\ 

\bottomrule
\end{tabular}
\vspace{-20pt}
}
\end{table}

\subsection {Sensitivity to Off-Chip Memory Bandwidth}
Table \ref{tab:bert_bandwidth} shows the effect of varying bandwidth on the latency of the BERT-Large model. 
We simulate different bandwidths by adjusting the amount of data moved from/to off-chip.
For example, we halve the data transferred from off-chip and pad the remainder on-chip to simulate 2X BW.
The first two data columns represent the theoretical minimum achievable latency if the bandwidth were infinite and there were no compute setup overheads, and if the computational resources were infinite.
Increasing bandwidth does not yield significant benefits, suggesting that the current use of bandwidth is already highly efficient.
Also, it shows that the current execution achieves 78.6\% utilization of the peak bandwidth.

\begin{table}[t]
\centering
\caption{BERT-Large Sweep Bandwidth Analysis, Sequence Length=384, Batch=8}
\resizebox{\linewidth}{!}{
\label{tab:bert_bandwidth}
\begin{tabular}
{p{2cm}p{1.8cm}p{1.3cm}p{0.45cm}p{0.4cm}p{0.45cm}p{0.45cm}}
\toprule
\textbf{Scenario} & \textbf{Infinite BW \newline \& No setup} &  \textbf{Infinite compute} & \textbf{0.5X BW} & \textbf{1X BW} & \textbf{2X BW} & \textbf{3X BW} \\
\midrule
\textbf{Latency (ms)} & 311 &  349 & 704 & 444 & 387 & 372 \\
\hline
\textbf{Speedup} & 1.43 & 1.27 & 0.63 & 1 & 1.15 & 1.19 \\
\bottomrule
\end{tabular}
}
\end{table}



\section{Related Work}
The dataflow architecture community has brought us significant inspiration. Decoupled access/execution architecture \cite{dataflow-decoupledaccess} decouples operand access and execution through two separate instruction streams. 
Streaming dataflow \cite{stream-dataflow, MozartReuse, DSAGEN} further explores this idea by abstracting computation as a CGRA-mapped dataflow graph and data movement
with streams and barriers.
Vector machines like Cray \cite{vector-machine-1, vector-machine-2, vector-machine-3, vector-machine-4} use vector abstraction to implicitly perform chaining in microarchitectures by making one vector directly feed into the next dependent vector. 
dMTCGRA \cite{dmt-cgra, mt-cgra-1, mt-cgra-2} allows programmers to explicitly write inter-thread communications in CGRA-mapped CUDA kernels without performing redundant memory loads. 
DySER \cite{dataflow-dyser} integrates a circuit-switched network of stateless FUs into the execution stage of a processor pipeline.
Triggered instruction \cite{dataflow-triggered} removes the program counter and integrates the architectural state registers into the FUs. 
Stream processors \cite{programmable-stream-processor, Stream-proga, imagine-stream} use streams to manage communication between stream registers and off-chip memory, and use customized kernel instructions to launch execution.

Streaming is a key concept in many FPGA designs and tools \cite{fpga-cpu-fpga, fpga-sextans, fpga-streamGCN, fpga-rapidstream2, fpga-levelst, fpga-rapidstream, fpga-stream-hls, fpga-tapa, 
dynamic-hls2,fpga-dynamichls, fpga-eth, fpga-eth2, stream-commomcase, stream-pigasus, stream-pigasus2}. 
Concurrently developed with our work, InTAR \cite{he2025intar} also targets dynamic resource allocation on FPGAs. Unlike overlays, it generates a dedicated bitstream per DNN model for aggressive pruning of redundant circuits. It uses a reconfigurable PE array with statically embedded partitioning configurations to enable dynamic layer pipelining, but lacks fine-grained memory bandwidth allocation.


\section{Conclusion}

This paper proposes the reconfigurable stream network architecture, introducing a network abstraction at the ISA level for DNN accelerators.
This novel abstraction aligns naturally with the structure of DNN workloads and the heterogeneous organization of modern hardware resources.
We envision future work in compiler support and software tooling to fully unlock the flexibility and programmability of RSN overlays.
We also believe that RSN has the potential to be applied beyond DNNs to other streaming-intensive domains, such as scientific computing.

%

\begin{acks}

We thank the reviewers from PPoPP'25 and ISCA'25 for their valuable feedback. We thank Sitao Huang, Stephen Neuendorffer, Tony Nowatzki, Jian Weng, and Zifan He for their insightful comments on the paper;
Gagandeep Singh and Joseph Melber for help with presentations at ISCA;
Andrew Boutros and Stephen More for teaching us the details of Intel FPGA NPU; Florent Werbrouck for his prompt and expert support on the AMD forums; and Marci Baun for editing the paper. We also thank our A-level cohort at CMU for the technical discussions and random thoughts along the way.

This work was partially supported by the NSF Award 2211557, CDSC industry partners (https://cdsc.ucla.edu/partners/), and the Intel/VMware Crossroads 3D-FPGA Academic Research Center. We acknowledge AMD’s hardware donation under the HACC Program.

\end{acks}



\appendix
\section{Artifact Appendix}

\subsection{Abstract}

This artifact provides the implementation of the RSN-XNN on an AMD VCK190 FPGA board for running the first encoder layer of \textbf{BERT-Large} (sequence length=512, batch size=6). It includes a bootable SD card image (\texttt{sd\_card.img}), input, weight and reference data (\texttt{python\_gold.zip}), and an optional container (\texttt{isca25-rsn.sif}) for regenerating the dataset. Following the instructions, functional correctness can be verified and final inference latency can be reproduced for the specified BERT configuration in Table~\ref{table:mm_operations}.
In addition, resource utilization and area overhead in Table~\ref{table:overhead} can be examined through the report generated during the bitstream compilation.

\subsection{Artifact check-list}
{\small
\begin{itemize}
  \item {\bf Algorithm:} BERT-Large 1st Encoder, Sequence Length = 512, Batch = 6, FP32.
  \item {\bf How much time is needed to prepare workflow?} 1-2 hours to setup VCK190 board. 4 hours to download and install Vitis.
  \item {\bf How much time is needed to complete experiments?} 20 hours. 
  \item {\bf Publicly available?} Yes.
  \item {\bf Code licenses:} AGPL-3.0 license.
  \item {\bf Archived:} \url{https://doi.org//10.5281/zenodo.15103085}.
\end{itemize}
}

\subsection{Description}

\subsubsection{How to access}
\begin{itemize}
  \item \textbf{GitHub:} \url{https://github.com/ChengyueWang/ISCA25-Stream-Network-Arch}.
  \item \textbf{Zenodo:} \url{https://doi.org/10.5281/zenodo.15102698}.
\end{itemize}

\subsubsection{Hardware dependencies}
\begin{itemize}
  \item \textbf{VCK190 Evaluation Kit}.
  \item \textbf{Linux server for bitstream generation:} Minimum 32 CPU cores, 64 GB RAM, 300 GB disk.
  \item \textbf{Local machine (Linux, Mac, or Windows):} Supports USB-UART and Ethernet connections.
\end{itemize}

\subsubsection{Software dependencies}
\begin{itemize}
  \item \textbf{AMD Vitis™ Unified Software Platform 2024.1}.
  \item \textbf{Vitis\_Libraries}.
  \item \textbf{Versal common image 2024.1}.
  \item \textbf{Petalinux 2024.1}.
  \item \textbf{Apptainer} (Optional, for containerized build of the dataset).
\end{itemize}

\subsubsection{Datasets}
\begin{itemize}
  \item \textbf{\texttt{python\_gold.zip}:} Contains input, model weight, and reference output for BERT-Large.  
\end{itemize}

\subsection{Installation}

Follow the instructions in \texttt{README.md} on GitHub to install the required software. Set the environment variables in \texttt{env.sh} for Vitis, PetaLinux, and the Versal common image. If desired, use \texttt{isca25-rsn.sif} with Apptainer to regenerate the dataset.

\subsection{Experiment workflow}

Precompiled \texttt{sd\_card.img} and \texttt{python\_gold.zip} are provided on Zenodo, so (1) and (2) can be skipped if you do not wish to regenerate the dataset or recompile the hardware bitstream. 
To reproduce the results from scratch:
\begin{enumerate}
  \item \textbf{Generate BERT data:} Run \texttt{./bert.sh} to produce input, weight, and output data under the folder \texttt{python\_gold/}.
  \item \textbf{Compile hardware bitstream:} Run \texttt{./fpga.sh} or submit \texttt{fpga-submit.batch} via Slurm. This will compile RSN-XNN and create \texttt{sd\_card.img}.
  \item \textbf{Deploy to VCK190:} Flash \texttt{sd\_card.img} to a microSD card and insert it into the VCK190 board.
  \item \textbf{Run inference:} Copy \texttt{python\_gold.zip} to the board, unzip it, and run \texttt{./run\_script.sh} to measure performance and verify correctness.
\end{enumerate}

\subsection{Evaluation and expected results}

The expected on-board inference time is 17.98 ms. Functional correctness should be confirmed by comparing the output, segment by segment, against the reference output provided in \texttt{python\_gold.zip}.
The total resource utilization for RSN-XNN will be reported in:
\path{RSN-XNN-FPGA/build/gemm_32x32x32/x1/hw/_x/link/vivado/vpl/prj/prj.runs/impl_1/full_util_routed.rpt.}
The area overhead of the decoder unit will be reported in:
\path{RSN-XNN-FPGA/build/gemm_32x32x32/x1/hw/_x/reports/dma_hls.hw/hls_reports/dma_hls_csynth.rpt.}
Note that non-deterministic behavior in Vitis can occur due to multi-threading, resulting in slight variations in resource utilization across synthesis runs.

\subsection{Experiment customization}

Users can modify the host files to generate different instructions:

\begin{itemize}
  \item \textbf{Optimization options (Table~\ref{table:mm_operations})}: In \path{RSN-XNN-FPGA/design/host_app_src/gemm_aie_app.cpp}, different optimization styles can be selected by commenting or uncommenting different \texttt{generate\_instruction\_*} functions. For example: \begin{itemize}
      \item Using \texttt{generate\_instruction\_onelayer} corresponds to ``No Optimize'' column in Table~\ref{table:mm_operations}.
      \item Using \texttt{generate\_instruction\_query\_key\_value} enables a fused operation.
    \end{itemize}
Users can freely toggle these options to observe performance differences of different optimizations or disable layers that are not under examination.

\item \textbf{Batch size and sequence length (Fig.~\ref{fig:input512} and Table~\ref{table:comparison_hardware})}: 
Users can try different problem settings by replacing the host folder \texttt{host\_app\_src/} using folders in 
\path{RSN-XNN-FPGA/design/host_different_config/}.
Users can experiment with various batch sizes (as in Fig.~\ref{fig:input512}) and a sequence length of 384 (as in Table~\ref{table:comparison_hardware}). 
The dataset for a sequence length of 384 is generated with files 
\path{BERT_HuggingFace/10-store_1layer384.py} and 
\path{11-inputlen384.cpp}.
\end{itemize}

\end{document}